\documentclass[journal = jctcce,
layout = twocolumn,
manuscript = article]{achemso}
\pdfoutput=1

\setkeys{acs}{
	articletitle  = false,
	maxauthors    = 5,
	doi           = true,
	super         = true
}

\usepackage[fontsize=11pt]{scrextend}

\SectionNumbersOn

\usepackage[hyperindex,breaklinks,hidelinks,colorlinks,citecolor=blue]{hyperref}
\usepackage{amsmath, amsthm, amssymb}
\usepackage{float}
\newcommand{\doi}[1]{\href{http://dx.doi.org/#1}{\nolinkurl{#1}}}
\usepackage{graphicx}
\DeclareGraphicsExtensions{.png}
\usepackage{booktabs,siunitx}
\allowdisplaybreaks
\usepackage{multirow}
\usepackage{natmove}

\newcommand{\vt}[1]{\boldsymbol{\mathbf{#1}}}

\author{Gabriela B. Correa}
\affiliation{Chemical Engineering Program, Instituto Alberto Luiz Coimbra de P\'os-Gradua\c{c}\~{a}o e Pesquisa em Engenharia, Universidade Federal do Rio de Janeiro, Rio de Janeiro, RJ 21941-909, Brazil}

\author{J\'essica C. S. L. Maciel}
\affiliation{Chemical Engineering Department, Escola de Qu\'imica, Universidade Federal do Rio de Janeiro, Rio de Janeiro, RJ 21941-909, Brazil}

\author{Frederico W. Tavares}
\affiliation{Chemical Engineering Program, Instituto Alberto Luiz Coimbra de P\'os-Gradua\c{c}\~{a}o e Pesquisa em Engenharia, Universidade Federal do Rio de Janeiro, Rio de Janeiro, RJ 21941-909, Brazil}
\alsoaffiliation{Chemical Engineering Department, Escola de Qu\'imica, Universidade Federal do Rio de Janeiro, Rio de Janeiro, RJ 21941-909, Brazil}

\author{Charlles R. A. Abreu}
\email{abreu@eq.ufrj.br}
\affiliation{Chemical Engineering Department, Escola de Qu\'imica, Universidade Federal do Rio de Janeiro, Rio de Janeiro, RJ 21941-909, Brazil}

\title{A New Formulation for the Concerted Alchemical Calculation of van der Waals and Coulomb Components of Solvation Free Energies}


\begin{document}

\maketitle

\begin{abstract}
Alchemical free energy calculations via molecular dynamics have been widely used to obtain thermodynamic properties related to protein-ligand binding and solute-solvent interactions. 
Although soft-core modeling is the most common approach, the linear basis function (LBF) methodology [Naden, L. N.; et al. \textit{J. Chem. Theory Comput.} \textbf{2014}, 10 (3), 1128; \textbf{2015}, 11 (6), 2536] has emerged as a suitable alternative.
It overcomes the end-point singularity of the scaling method while maintaining essential advantages such as ease of implementation and high flexibility for postprocessing analysis.
In the present work, we propose a simple LBF variant and formulate an efficient protocol for evaluating van der Waals and Coulomb components of an alchemical transformation in tandem, in contrast to the prevalent sequential evaluation mode.
To validate our proposal, which results from a careful optimization study, we performed solvation free energy calculations and obtained octanol-water partition coefficients of small organic molecules.
Comparisons with results obtained via the sequential mode using either another LBF approach or the soft-core model attest to the effectiveness and correctness of our method. 
In addition, we show that a reaction field model with an infinite dielectric constant can provide very accurate hydration free energies when used instead of a lattice-sum method to model solute-solvent electrostatics.
\end{abstract}


\section{Introduction}
\label{sec:introduction}

In recent years, calculations of free energy differences have been performed in several investigations to provide values of important thermodynamic properties \cite{Hansen_2014}.
Binding free energies of protein-ligand systems, for example, are applied in the discovery of new drugs through the prediction of structural stability and enzymatic activity \cite{Chodera_2011, Mobley_2012, Mobley_2006}.
Solvation free energy calculations in turn provide information on solubility and partition coefficients \cite{Chandran_2018, Espinosa_2018, Bannan_2016, Garrido_2009, Marrero_2002, Fan_2020}.
These properties are widely studied via molecular simulation \cite{Spicher_2020, Paquet_2015, Ross_2015}, with which the microscopic behavior of a system can be investigated.

Using molecular dynamics (MD), we can simulate alchemical transformations \cite{Hansen_2014, Chodera_2011}, meaning that the free energy difference between two states is computed using a nonphysical pathway.
This is possible because free energy is a state function, i.e., it depends only on the initial and final states of a process.
Intermediate states often connect the two extremes of the process so that the alchemical transformation occurs gradually.
Several such states are usually required to continuously cover the relevant region of phase space.
The required number of such states is essentially dependent on the choice of the alchemical pathway.

The states are defined by different values of a coupling parameter $\lambda$, which controls the strength of the potential energy of interaction between particles of interest.
For example, in calculating the solvation free energy, the initial state is a solute in a vacuum, and the final state is the same solute in a solvent medium.
The system potential energy can depend on $\lambda$ in such a way that the solute-solvent interaction is nullified when $\lambda = 0$ and fully activated when $\lambda = 1$ \cite{Khanna_2020}.
Different formulations of the potential energy as a function of $\lambda$ have been created, resulting in distinct alchemical pathways.

The scaling method \cite{Pitera_2002} consists of multiplying the original potential of the alchemically affected interactions by a $\lambda$-dependent switching function.
However, this method can lead to imprecise free energies in systems involving Lennard-Jones (LJ) interactions due to an end-point singularity.
The LJ potential goes to infinity as the distance between atoms approaches zero, which becomes an issue when $\lambda$ changes from zero to a small finite value.
To solve this shortcoming, \citet{Beutler_1994} proposed their well-known soft-core approach.

The soft-core model is widely applied in alchemical calculations \cite{Barbosa_2021, Lee_2020, Mihalovits_2020, Mermelstein_2018} and has been implemented in MD software packages such as GROMACS \cite{Hess_2008}, CHARMM \cite{Brooks_2009}, and AMBER \cite{Case_2005}.
This approach provides a smooth free energy profile as a function of $\lambda$ and yields a low-variance estimate \cite{Pham_2011, *Pham_2012}.
However, an important disadvantage stems from the modification done in the mathematical expressions of the LJ and Coulomb potentials to include $\lambda$.
It involves changes in the inner force loop of an MD code \cite{Naden_2014,Naden_2015}, which is usually highly optimized in terms of specialized hardware utilization.
Moreover, multiple-state free energy estimators, such as the multistate Bennett acceptance ratio (MBAR) method \cite{Shirts_2008}, require computation of the potential energy of each sampled configuration at all $\lambda$ states under consideration in the analysis.
With a soft-core potential, we must do such a calculation on the fly, which also demands changes in the inner loop of the MD code.
In addition, careful planning becomes necessary given the difficulty of including unanticipated states in the MBAR analysis.

In the scaling method, on the other hand, the LJ and Coulomb potentials remain unchanged as we multiply them by their respective switching functions.
Hence, despite the end-point singularity issue, it has considerable advantages over the soft-core method.
For example, we can defer the computation of potential energies at the multiple $\lambda$ states to a postprocessing stage \cite{Buelens_2012, Naden_2014}, and then we can apply the MBAR method to estimate the free energies and other properties at either sampled or unsampled $\lambda$ states via reweighting \cite{Kong_2003, Shirts_2008}.
Such unsampled states can even be defined using switching functions other than those employed in the simulations.
This opens the possibility of searching for optimal switching functions according to some criterion of choice without performing new simulations.

Recently, Naden et al. \cite{Naden_2014, Naden_2015} proposed a general framework for the scaling approach that they called the linear basis function (LBF) model.
They managed to avoid the end-point singularity by softening short-range interactions \cite{Buelens_2012} in the first step of the alchemical transformation and then in a later step applying a correction to restore the original LJ potential.

Here we propose an LBF model that is mathematically simpler and involves fewer parameters than the original one.
We provide a set of switching function parameters obtained by optimizing the calculation of hydration free energies of selected solutes.
In contrast to most current methods \cite{Abreu_2020a, Barbosa_2021}, our approach aims to provide concerted alchemical transformations, in which the total free energy difference, including both the van der Waals and Coulomb contributions, is obtained after a single sweep of $\lambda$ values from $0$ to $1$.

Concerted alchemical transformations are suitable for advanced sampling schemes such as $\lambda$ dynamics \cite{Kong_1996, Raman_2020} and adaptive biasing \cite{Babin_2008, Hayes_2017}.
Furthermore, they are more convenient than sequential evaluation approaches, where each step is associated with different sampling requirements and statistical errors \cite{Lee_2020}.
Other authors evaluated transformations with simultaneous changes in the LJ and Coulomb potentials, applying either a modified soft-core model \cite{Lee_2020} or a capping function that acts directly on the total alchemical potential energy rather than on each alchemically affected pair interaction \cite{Pal_2019, Wu_2021, Azimi_2022}.
In contrast, our proposal relies on the LBF approach, thus bringing about all of the benefits mentioned earlier.

To test the effectiveness and robustness of our LBF method, we performed MD simulations to calculate the solvation free energies of organic molecules in both water and 1-octanol. 
This allowed us to compute their octanol-water partition coefficients and compare results across different methods and with experimental and simulation data from the literature.

In addition, while computing hydration free energies, we analyzed the use of different pairwise electrostatic approaches \cite{Tironi_1995, Kubincova_2020, Fennell_2006a} to model solute-solvent interactions in place of the particle mesh Ewald (PME) method \cite{Darden_1993, Essmann_1995}.
We observed that a reaction field model \cite{Tironi_1995, Kubincova_2020} could provide highly accurate results if the dielectric constant of the medium is considered to be infinite, as in a conductor.
The results deviate slightly from PME-based ones, and if necessary, they can be corrected with high precision through a simple perturbation method \cite{Zwanzig_1954}.
Along with the proposed LBF formulation, this observation is an important contribution to the field of alchemical free energy calculation.

The remainder of this article is organized as follows.
In section~\ref{sec:methodology}, we present the applied methodology, which
includes the proposed LBF model and the schemes we used for obtaining optimized parameters.
Results are presented in section~\ref{sec:results}.
Finally, we summarize our conclusions in section~\ref{sec:conclusion}.

\section{Methodology}
\label{sec:methodology}

\subsection{Linear Basis Function Approach for Alchemical Transformations}
\label{sec:generalLBF}

In the LBF approach, formalized in a series of publications by Naden et al. \cite{Naden_2014, Naden_2015}, the total potential between two alchemically interacting particles at a distance $r$ is written as
\begin{equation}
\label{eq:total_alchemical_LBF}
u(r, \lambda) = \sum_k h_k(\lambda) u_k(r)
\end{equation}
where $u_k(r)$ is the potential basis function of type $k$ and $h_k(\lambda)$
is its corresponding switching function.
The choice of basis functions can be made so as to avoid the end-point singularity described in the Introduction.
Additionally, a judicious choice of switching functions can result in an effective alchemical protocol.

The original LBF model relies on a strategy proposed by \citet{Buelens_2012} (BG) to eliminate the end-point singularity.
It consists of forcing the LJ potential to detour from its path to infinity as $r$ goes to zero, but only at an energy value high enough for the change to have a negligible impact on the sampling probabilities under usual thermodynamic conditions.
In this way, the modified and original models, which \citet{Buelens_2012} called the \textit{capped} and \textit{native} pair potentials, respectively, yield free energy values indistinguishable from each other.
However, \citet{Naden_2014} observed that enforcing the detour at lower potential energies produces free energy results with smaller variances but at the cost of introducing non-negligible deviations from the native-model results.
As a solution, they added a new step to the free energy calculation to account for the \textit{residual} (or decapping) contribution, defined as the difference between the native and capped potentials.

Naden et al. \cite{Naden_2014, Naden_2015} employed a basis with four potential functions.
Besides using the Coulomb interaction for charged atoms, they performed a Weeks-Chandler-Andersen (WCA) decomposition \cite{Weeks_1971} of the LJ potential into repulsive and attractive contributions, followed by the BG-type splitting of the repulsive part.
Their version of the capped repulsive potential requires some constants to be calculated for each individual set of LJ parameters.
They defined the switching function corresponding to each contribution so as to reduce the total variance of the alchemical calculations.
The authors took advantage of the reweighting capability of the LBF approach in order to carry out a very efficient optimization procedure.
They ended up recommending the trivial linear coupling for all contributions except the capped repulsive potential, for which they suggested a fourth-degree polynomial with tuned coefficients.
They also tested different orderings for the coupling of van der Waals and electrostatic interactions. One of the best performances was achieved by switching on the attractive and the capped repulsive potentials simultaneously at first, followed by the residual potential and finally by the electrostatic contribution.

In the present work, we propose a simpler formulation.
Instead of using the WCA decomposition, we apply a BG-type capping to the full LJ model.
Our LBF potential for the alchemical interactions is written as
\begin{equation}
u(r, \lambda) = h_\textsc{c}(\lambda) u_\textsc{c}(r) + h_\textsc{r}(\lambda) u_\textsc{r}(r) + h_\textsc{e}(\lambda) u_\textsc{e}(r)
\label{eq:LBFinitial}
\end{equation}
where $u_\textsc{c}(r)$, $u_\textsc{r}(r)$, and $u_\textsc{e}(r)$ are the capped, residual, and electrostatic pair potentials, respectively.

The LJ potential is expressed as $u_\textsc{lj}(r) = \epsilon \nu_\textsc{lj} (r/\sigma)$, where $\sigma$ and $\epsilon$ are the usual LJ parameters \cite{Harrison_2018} and $\nu_\textsc{lj}(x)$ is a reduced potential given by
\begin{equation}
\nu_\textsc{lj}(x) = 4(x^{-12} - x^{-6})
\label{eq:LJpotential}
\end{equation}
We developed a simple version of the BG potential, in which a reduced capping function $\nu_\mathrm{cap}(x)$ replaces $\nu_\textsc{lj}(x)$ for all $x < 1$.
Therefore, $u_\textsc{bg}(r) = \epsilon \nu_\textsc{bg} (r/\sigma)$, where
\begin{align}
\label{eq:BGpotential}
\nu_\textsc{bg}(x) =\left\{
\begin{array}{cl}
\nu_\mathrm{cap}(x), & \quad \text{for} \quad x < 1  \\
\nu_\textsc{lj}(x), & \quad \text{for} \quad x \geq 1
\end{array}
\right.
\end{align}
The capping function is defined as a polynomial of degree $m$ satisfying the conditions $\nu_\mathrm{cap}^{(n)}(0) = 0$ for $n = 1, \dots, m_0$ and $\nu_\mathrm{cap}^{(n)}(1) = \nu_\textsc{lj}^{(n)}(1)$ for $n = 0, \dots, m_1$, where $m_0 + m_1 = m$ and $f^{(n)}(x)$ is the $n$th order derivative of $f(x)$.
These conditions, which completely determine the polynomial coefficients, guarantee that $\nu_\textsc{bg}(x)$ is continuous and smooth at $x = 1$ and singularity-free at $x = 0$.
The polynomial that fulfills these conditions with $m_0 = m_1 = 3$ is
\begin{equation}
\label{eq:BGp}
\nu_\mathrm{cap}(x) = -868x^6 + \frac{10944}{5} x^5 -1440x^4 + \frac{596}{5}
\end{equation}

\begin{figure}[htbp!]
	\centering
	\includegraphics[scale=1.0]{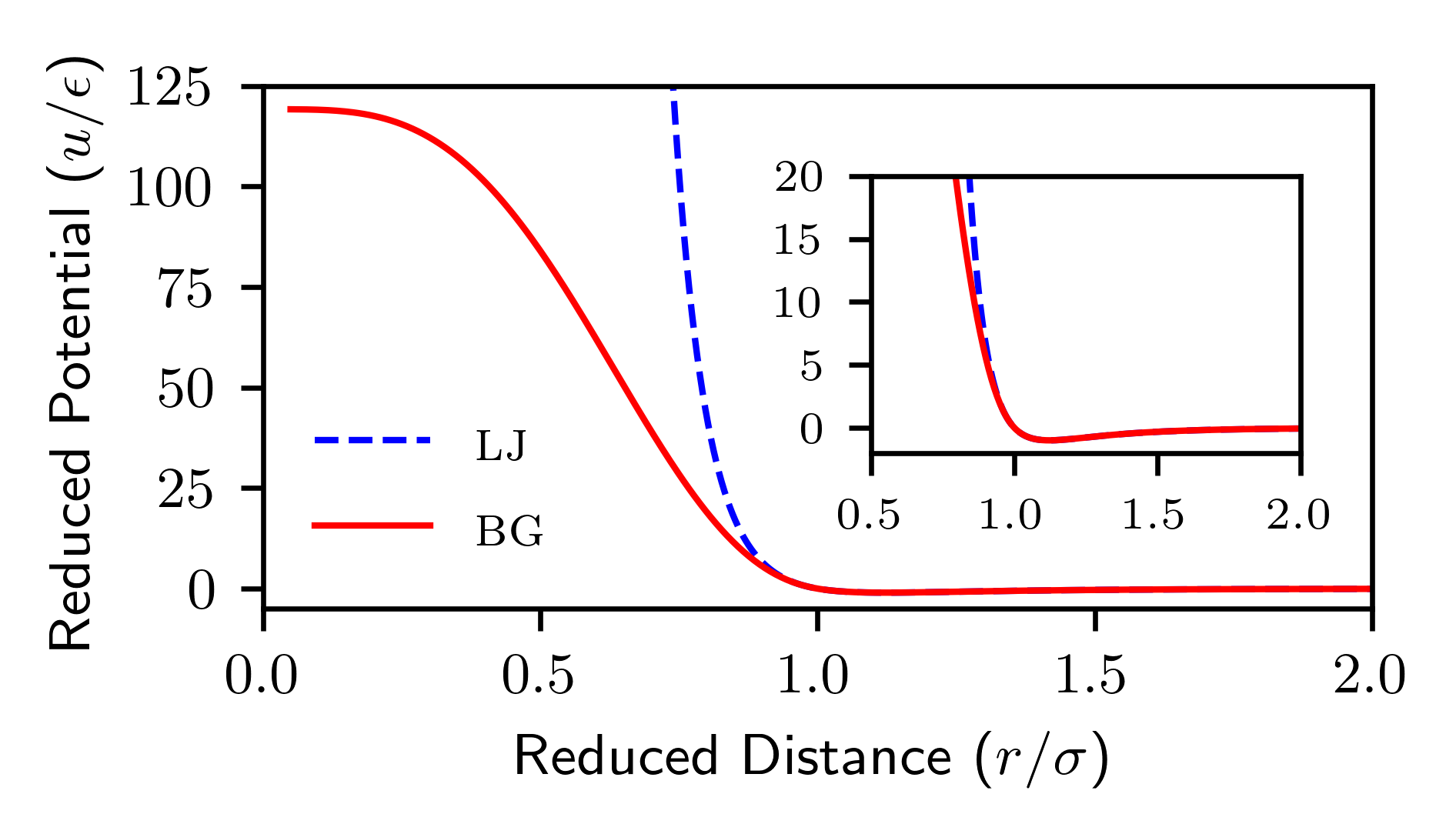}
	\caption{
		Lennard-Jones potential (blue dashed line) and the Buelens-Grubm\"uler-type modification \cite{Buelens_2012} proposed in this work (red solid line), which is aimed at softening short-range interactions and eliminating end-point singularities in solvation free energy calculations.
	}
	\label{fig:BG}
\end{figure}

Therefore, rather than diverging to infinity, the potential energy approaches $119.2 \epsilon$ as $r \rightarrow 0$.
As an example, this value is close to $42~\mathrm{kJ/mol}$ in the case of the LJ interaction between oxygen atoms in the TIP3P water model \cite{Jorgensen_1983}.
Figure~\ref{fig:BG} shows a comparison between the native and capped potentials in their reduced forms, given by eqs~\eqref{eq:LJpotential} and \eqref{eq:BGpotential}, respectively.
The main advantage of the capped potential defined above, in contrast to those used elsewhere \cite{Naden_2014, Naden_2015, Buelens_2012}, is its simplicity.
By changing $m_0$ it is also possible to alter the potential energy value at $r=0$.
Nevertheless, the reduced potential in eq~\eqref{eq:BGp} succeeded in all studies performed here.

\begin{figure}[htbp!]
	\centering
	\includegraphics[scale=1.0]{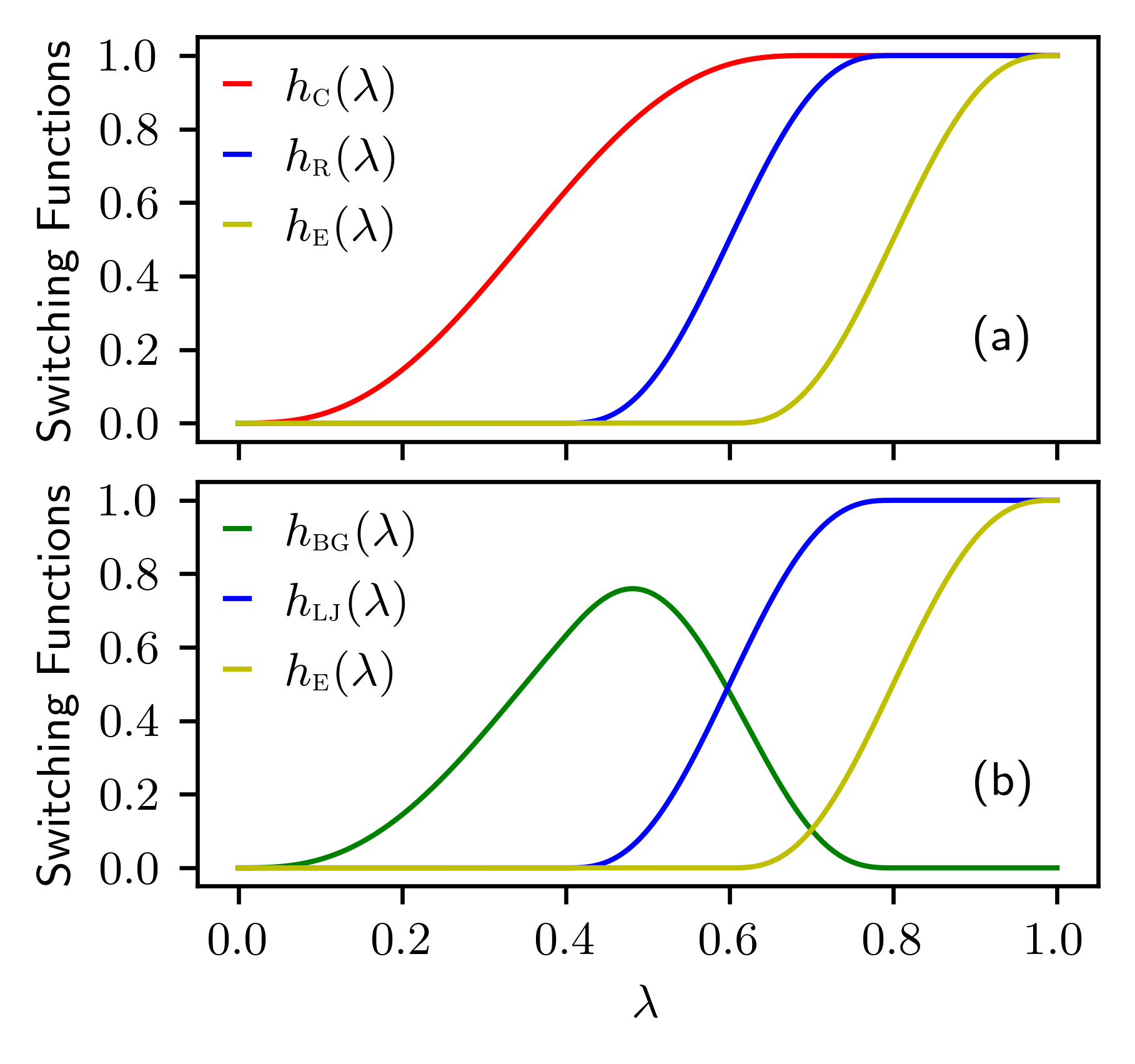}
	\caption{Alternative representations via (a) eq~\eqref{eq:LBFinitial} and (b) eq~\eqref{eq:alternative coupling scheme} of the smooth switching functions proposed in this work for a single-step, concerted coupling of the capped, residual, and electrostatic potentials. In both cases, eq~\eqref{eq:sfnova} was used with parameters $\lambda_\textsc{c}^\text{s} = 0.0$, $\lambda_\textsc{c}^\text{f} = 0.7$, $\lambda_\textsc{r}^\text{s} = 0.4$, $\lambda_\textsc{r}^\text{f} = 0.8$, $\lambda_\textsc{e}^\text{s} = 0.6$, and $\lambda_\textsc{e}^\text{f} = 1.0$.
}
	\label{fig:SF_smooth}
\end{figure}

In summary, with the definitions above for the capped and residual potentials, we can rewrite eq~\eqref{eq:LBFinitial} as
\begin{multline}
\label{eq:coupling scheme}
u(r, \lambda) =  h_\textsc{c}(\lambda) u_\textsc{bg}(r) + \\
h_\textsc{r}(\lambda) [u_\textsc{lj}(r) - u_\textsc{bg}(r)] +
h_\textsc{e}(\lambda) u_\textsc{e}(r)
\end{multline}
A sensible choice of switching functions $h_\textsc{c}$, $h_\textsc{r}$, and $h_\textsc{e}$ can reduce the number of $\lambda$ states required to obtain free energy estimates with a desirable precision.
Here we propose a family of smooth switching functions designed to allow the coupling of all basis potentials in a single sweep of $\lambda$ values from $0$ to $1$.
They are defined as
\begin{equation}
h_k(\lambda) = S \left( \frac{\lambda - \lambda_k^\text{s}}{\lambda_k^\text{f} - \lambda_k^\text{s}} \right)
\label{eq:sfnova}
\end{equation}
where $\lambda_k^\text{s}$ and $\lambda_k^\text{f}$ are the starting and finishing points of the actual switching of the $k$th basis potential, respectively, and $S(x)$ is a twice-differentiable smoothstep function given by \cite{Lee_2020}
\begin{align}
S(x) =\left\{
\begin{array}{lll}
0, & \text{for} & x < 0  \\
6x^5 - 15x^4 + 10x^3, & \text{for} & 0 \leq x \leq 1\\
1, & \text{for} & x > 1  \\
\end{array}
\right. 
\label{eq:SFsmooth}
\end{align}
Thus, multiple partially coupled alchemical potentials can coexist at the same $\lambda$ value, such as in the example shown in Figure~\ref{fig:SF_smooth}(a).
In other words, a particular coupling stage can start before the previous one has finished.

An interesting reinterpretation of the coupling scheme just described can be observed in Figure~\ref{fig:SF_smooth}(b).
It stems from rewriting eq~\eqref{eq:coupling scheme} as
\begin{equation}
\label{eq:alternative coupling scheme}
u(r, \lambda) =  h_\textsc{bg}(\lambda) u_\textsc{bg}(r) + h_\textsc{lj}(\lambda) u_\textsc{lj}(r) + h_\textsc{e}(\lambda) u_\textsc{e}(r)
\end{equation}
where $h_\textsc{bg} = h_\textsc{c} - h_\textsc{r}$ and $h_\textsc{lj} = h_\textsc{r}$.
Hence, at the smallest $\lambda$ values, only the capped potential is coupled to the system, thus preventing the end-point singularity.
Then, when it is safe to do so, the native LJ potential progressively replaces its capped counterpart.

\begin{figure}[htbp!]
	\centering
	\includegraphics[scale=1.0]{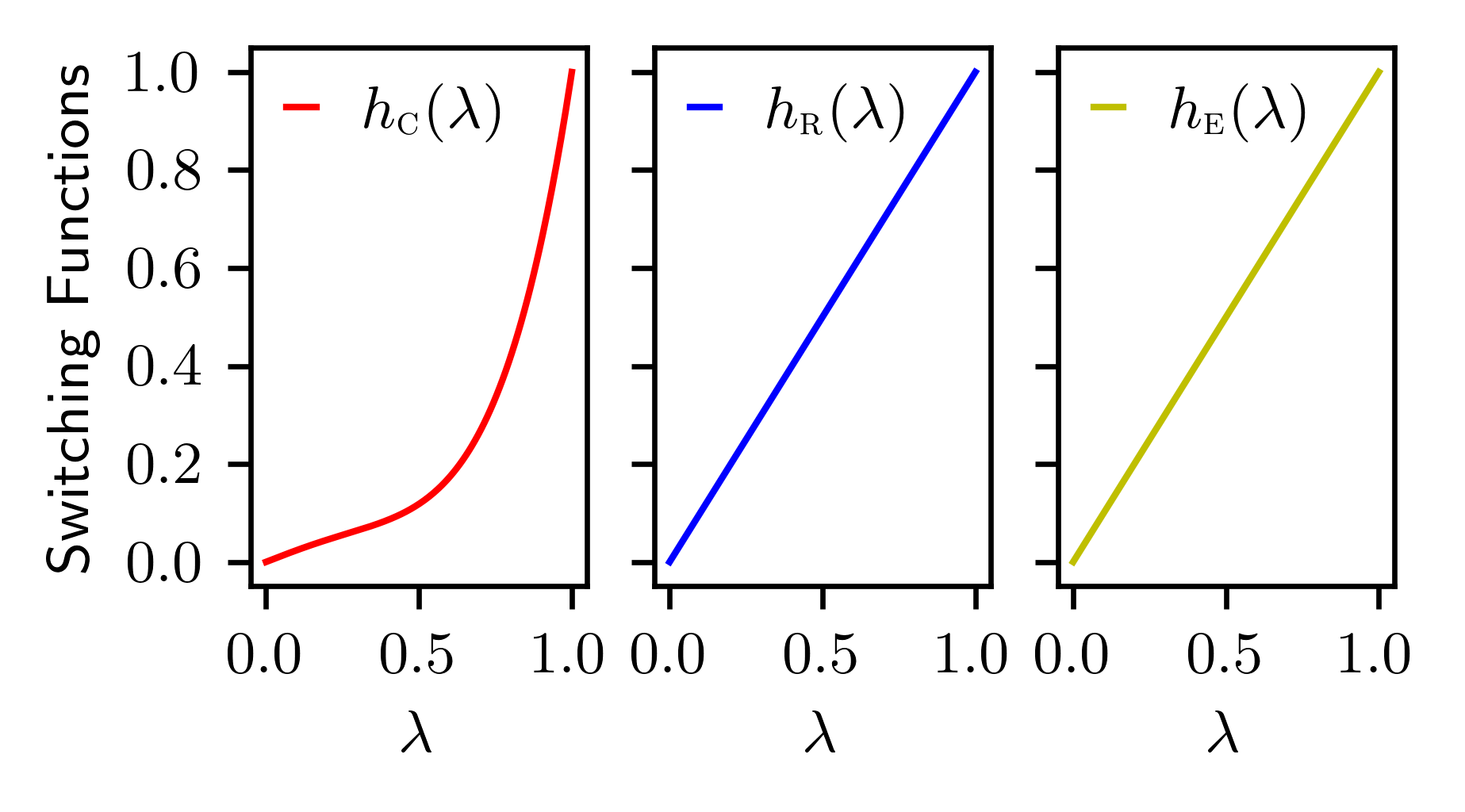}
	\caption{Quartic and linear switching functions used for multiple-step sequential coupling of the capped, residual, and electrostatic potentials.}
	\label{fig:SF_shirts_scheme}
\end{figure}

For comparison, we also apply the formulation of Naden et al. \cite{Naden_2014, Naden_2015} to our particular splitting of the LJ potential.
It consists of sequential coupling of the capped, residual, and electrostatic potentials separately.
For this, we resort to a quartic switching function \cite{Naden_2014},
\begin{equation}
	P_k(\lambda) = A_k \lambda^4 + B_k \lambda^3 + C_k \lambda^2 + (1-A_k-B_k-C_k) \lambda
	\label{eq:shirts1}
\end{equation}
In the initial step, we make $h_\textsc{c} = P_\textsc{c}$ while keeping $h_\textsc{r} = h_\textsc{e} = 0$.
In the middle step, we make $h_\textsc{c} = 1$, $h_\textsc{r} = P_\textsc{r}$, and $h_\textsc{e} = 0$.
Then, in the final step, we keep $h_\textsc{c} = h_\textsc{r} = 1$ and make $h_\textsc{e} = P_\textsc{e}$.
This procedure is illustrated in Figure~\ref{fig:SF_shirts_scheme}, whose each panel depicts the nonconstant switching function used in each step.
Applying linear scaling (i.e., $A_k=B_k=C_k=0$) for both the residual and electrostatic potentials, as shown in the figure, was the recommendation of Naden et al. \cite{Naden_2014, Naden_2015}, who also applied it for the attractive part of the WCA potential.
For the capped repulsive part, they recommended using eq~\eqref{eq:shirts1} with $A = 1.62$, $B = -0.889$, and $C = 0.0255$.
Here we employ these values as an initial guess for obtaining an optimal switching function (see section~\ref{sec:optimizations}) for our capped LJ potential.

\subsection{Alchemical Coupling of Solute-Solvent Interactions}
\label{sec:electrostatic}

When using our LBF model for computing solvation free energies, we adopt the so-called coupling route \cite{Abreu_2020a, Khanna_2020}, in which only solute-solvent interactions undergo alchemical transformation.
Hence, the total potential energy of a system as a function of the particle coordinates $\vt r$ and coupling parameter $\lambda$ is given by
\begin{equation}
U(\vt r, \lambda) = U_\text{b}(\vt r) + U_\text{nb}^\textsc{aa+ss}(\vt r) + \sum_k h_k(\lambda) U_k^\textsc{as}(\vt r)
\label{eq:LBFtotal}
\end{equation}
where the letters A and S refer to the solute and the solvent, respectively.
The potential $U_\text{b}$ includes all bonded interactions (bond stretching, bending, torsion, and so on) in the system, while $U_\text{nb}^\textsc{aa+ss}$ refers to the solute-solute and solvent-solvent nonbonded interactions, which are not affected by the alchemical transformation.
We employed a switched and truncated LJ model and PME electrostatic calculations to evaluate $U_\text{nb}^\textsc{aa+ss}$.
The alchemically affected potentials $U_k^\textsc{as}$ are intermolecular interactions between solute and solvent.
The parts associated with van der Waals interactions can be represented as
\begin{equation}
U_k^\textsc{as}(\vt r) = \sum_{i \in \textsc{a}} \sum_{j \in \textsc{s}} S \left( \tfrac{r_c - r_{ij}}{r_c - r_s} \right) \epsilon_{ij} \nu_k\left(\tfrac{r_{ij}}{\sigma_{ij}}\right)
\end{equation}
where $k$ is either LJ or BG, $r_c$ is the cutoff distance, $r_s$ is a switching distance, and $S(x)$ is given by eq~\eqref{eq:SFsmooth}.
For every pair formed by a solute atom $i$ and a solvent atom $j$ located at Cartesian positions $\vt r_i$ and $\vt r_j$, respectively, we have $r_{ij} = \|\vt r_i - \vt r_j\|$, while $\epsilon_{ij}$ and $\sigma_{ij}$ are the pair-specific LJ parameters, usually obtained from their atom-specific counterparts via mixing rules \cite{Harrison_2018}.
It is important to note the presence of the reduced pair potentials $\nu_\textsc{lj}(x)$ and $\nu_\textsc{bg}(x)$, defined in eqs~\eqref{eq:LJpotential} and \eqref{eq:BGpotential}, respectively.

The electrostatic part of the solute-solvent interaction is computed using a reaction field model \cite{Tironi_1995, Kubincova_2020} considering a perfectly conducting medium whose dielectric constant is infinite.
In this case, the solute-solvent electrostatic potential energy is given by
\begin{equation}
U_\textsc{crf}^\textsc{as}(\vt r) = \sum_{i \in \textsc{a}} \sum_{j \in \textsc{s}} S \left( \tfrac{r_c - r_{ij}}{r_c - r_s} \right) \frac{q_i q_j}{4 \pi \epsilon_0 r_c} \nu_\textsc{crf} \left(\tfrac{r_{ij}}{r_c}\right)
\label{eq:crf_complete}
\end{equation}
where CRF stands for conductor-like reaction field, $q_i$ and $q_j$ are the electric charges of atoms $i$ and $j$, respectively, and $\epsilon_0$ is the permittivity of free space.
The reduced potential is
\begin{equation}
\nu_\textsc{crf} (x) = \frac{1}{x} + \frac{x^2 - 3}{2}
\label{eq:CRF_reduce}
\end{equation}
which satisfies the conditions $\nu_\textsc{crf}(1) = \nu'_\textsc{crf}(1) = 0$, thus ensuring continuity of both the potential and its resulting force at $r_{ij} = r_{c}$.
Nevertheless, a smooth switching from $r_s$ to $r_c$ was applied so that the second-order derivative was continuous as well.
The distance between atoms throughout the simulations will never be zero, avoiding a singularity in the electrostatic potential. 
This necessarily occurs because the electrostatic potential is only turned on when the LJ interactions are at least partially coupled, creating a repulsive layer on the particles and thus preventing opposite charges from collapsing by attraction.

A cutoff pairwise treatment is applied for the solute-solvent electrostatics due to the ease of implementation and computational efficiency.
As we will show in section~\ref{sec:solute-solvent electrostatic interactions} through empirical tests, simulations using $U_\textsc{crf}^\textsc{as}$ provide free energy results that are often very close to those obtained using $U_\textsc{pme}^\textsc{as}$ instead.
It is in fact the best choice for this purpose among a set of tested pairwise electrostatic models.
Furthermore, one can readily apply a correction by sampling the difference $U_\textsc{pme}^\textsc{as} - U_\textsc{crf}^\textsc{as}$ at regular intervals during the simulation at the final $\lambda$ state and then performing a free energy perturbation (FEP) analysis \cite{Zwanzig_1954} afterward.
Hence, the required correction is estimated by means of the identity
\begin{equation}
{\left[G_\textsc{pme}^\textsc{as} - G_\textsc{crf}^\textsc{as}\right]}_\textsc{fep} = -k_\textsc{b} T \ln \left\langle e^{-\beta (U_\textsc{pme}^\textsc{as} - U_\textsc{crf}^\textsc{as})} \right\rangle_{\lambda = 1}
\end{equation}
where $k_\textsc{b}$ is the Boltzmann constant, $T$ is the system temperature, $\beta = (k_\textsc{b} T)^{-1}$, and $\langle \cdot 	\rangle_{\lambda = 1}$ denotes an ensemble average at the final $\lambda$ state.
We will also show in section~\ref{sec:solute-solvent electrostatic interactions} that FEP calculations are sufficiently precise for doing this type of correction, thus dispensing with the need for more elaborate methods \cite{Bennett_1976}.

\subsection{Optimization Schemes}
\label{sec:optimizations}

We solved different minimization problems to obtain optimal parameters for the switching functions defined in eqs~\eqref{eq:sfnova} and \eqref{eq:shirts1}.
We considered alternative objective functions, namely,
(i) a measure of local variations in the $\langle \partial U/ \partial \lambda \rangle$ versus $\lambda$ curve and
(ii) the asymptotic variance of the coupling free energy.

When the sampling effort of the MD simulations is uniformly spread over the whole range from $\lambda=0$ to $\lambda=1$, regions with free energy variations that are too large or too small will result in inefficient sampling due to, respectively, poor or excessive probability distribution overlaps between adjacent states.
Supposing that the amount of overlap is directly proportional to the free energy difference between two states \cite{Wu_2005a, *Wu_2005b},
an ideal free energy profile would be a straight line from $G(0)$ to $G(1)$, hence with constant $\langle \partial U/ \partial \lambda \rangle = \Delta G$.
Since most cases of interest involve a free energy barrier, another efficient profile would entail a smooth increase from $G(0)$ up to the barrier top, followed by a smooth decrease from there down to $G(1)$.
Although these ideal profiles are most often unrealizable for a given family of switching functions, we can look for a profile that minimizes some measure of local variations in the mean derivative $\langle \partial U/\partial \lambda \rangle = G'(\lambda)$.
Here we use a finite-difference approximation for the mean absolute second-order derivative, $\int_0^1 |G''(\lambda)| {\rm d} \lambda$, given by
\begin{equation}
L = \sum_{i=1}^{N_p} \left| G' \left( \tfrac{i}{N_p} \right) - G' \left( \tfrac{i-1}{N_p} \right) \right|
\label{eq:fobj1}
\end{equation}
where the $\lambda$ domain has been divided into $N_p$ equally sized intervals.

Let us now discuss variance minimization, which Naden et al. \cite{Naden_2014, Naden_2015} also applied to obtain optimal coefficients for eq~\eqref{eq:shirts1} when it was used as a switching function for their repulsive WCA potential.
As they did, we compute the free energy variance using the expression
\begin{equation}
\text{var}(\Delta G) = \int_0^1 \text{var} \left( \left\langle \frac{\partial U}{\partial \lambda} \right\rangle \right) d \lambda
\label{eq:varianceMETHODS}
\end{equation}  
which is related to the thermodynamic integration method \cite{Kirkwood_1935}. 
In this case, we evaluate the variance of the mean derivative $\langle \partial U/\partial \lambda \rangle$ and numerically integrate it with the $\lambda$ range.
The mean derivative $\langle \partial U/\partial \lambda \rangle$ can be obtained in a postprocessing stage by
\begin{equation}
\label{eq:derivative U}
\left\langle \frac{\partial U}{\partial \lambda} \right\rangle = \sum_k \left\langle U_k^\textsc{as} \right\rangle \frac{\partial h_k}{\partial \lambda}
\end{equation}
We apply the MBAR method to estimate both the value and the variance of the mean derivative at either sampled or unsampled states through reweighting \cite{Kong_2003, Shirts_2008}.
Most importantly, the LBF approach allows us to perform reweighting aiming at states defined for switching functions other than those employed in the simulations.
This possibility makes the optimization process very fast because evaluating the objective function for new parameter values does not require new MD simulations.

We define three optimization problems with the objective functions indicated in eqs~\eqref{eq:fobj1} and \eqref{eq:varianceMETHODS}.
They will be called schemes S1, S2, and Q, where S and Q stand for smooth and quartic switching functions, respectively.
They all require an initial guess of the switching function parameters, which we use as a reference system to perform MD simulations.
The number of simulated $\lambda$ states, which will be presented for each case in section~\ref{sec:Simulation Details}, must be sufficiently high to allow reweighting with appropriate precision.
The optimization routine is carried out using the Powell algorithm \cite{Powell_1964}, and eq~\eqref{eq:varianceMETHODS} is integrated through Gaussian quadrature.
We use the SciPy library \cite{Virtanen_2020} to apply both methods.

\subsubsection{Schemes S1 and S2}

Schemes S1 and S2, which rely on the smooth switching function defined in eq~\eqref{eq:sfnova}, differ only in the employed objective function.
The former entails using $L$ from eq~\eqref{eq:fobj1} as the objective function, while the latter entails using $\text{var}(\Delta G)$ from eq~\eqref{eq:varianceMETHODS}.
In both schemes, the reference system for carrying out the simulations was specified with the parameters values $\lambda_\textsc{c}^\text{s} = 0.0$, $\lambda_\textsc{c}^\text{f} = 0.7$, $\lambda_\textsc{r}^\text{s} = 0.4$, $\lambda_\textsc{r}^\text{f} = 0.8$, $\lambda_\textsc{e}^\text{s} = 0.6$, and $\lambda_\textsc{e}^\text{f} = 1.0$.
During the optimizations, the couplings of the capped and electrostatic potentials were forced to start at $\lambda=0$ and finish at $\lambda=1$, respectively, meaning that
$\lambda_\textsc{c}^\text{s}$ and $\lambda_\textsc{e}^\text{f}$ were excluded from the set of optimization variables.

Two types of optimizations were performed.
The first one involved four optimization variables, namely, $\lambda_\textsc{c}^\text{f}$, $\lambda_\textsc{r}^\text{s}$, $\lambda_\textsc{r}^\text{f}$, and $\lambda_\textsc{e}^\text{s}$, under the constraints $\lambda_\textsc{r}^\text{s} \leq \lambda_\textsc{c}^\text{f}$ and $\lambda_\textsc{e}^\text{s} \leq \lambda_\textsc{r}^\text{f}$,
so that each coupling stage can finish only if the subsequent one has already started.
The second type was a special case in which the residual and electrostatic potentials were coupled together, meaning that $\lambda_\textsc{r}^\text{f} = 1$ and $\lambda_\textsc{e}^\text{s} = \lambda_\textsc{r}^\text{s}$.
In this case, the only two optimization variables were $\lambda_\textsc{c}^\text{f}$ and $\lambda_\textsc{r}^\text{s}$, under the constraint $\lambda_\textsc{r}^\text{s} \leq \lambda_\textsc{c}^\text{f}$.

\subsubsection{Scheme Q}

Optimization procedures were applied independently to the coupling of the capped, residual, and electrostatic potentials.
For each case, the objective function of eq~\eqref{eq:varianceMETHODS} was employed with the quartic polynomial defined in eq~\eqref{eq:shirts1} as the switching function.
Thus, the optimal values of the parameters $A_k$, $B_k$, and $C_k$ were sought in each case.
The reference system for carrying out the simulations was specified based on ref~\citenum{Naden_2014}, with linear scaling for the residual and electrostatic potentials and parameters $A_\textsc{c} = 1.62$, $B_\textsc{c} = -0.889$, and $C_\textsc{c} = 0.0255$ for the capped potential.

\subsection{Octanol-Water Partition Coefficient}

The octanol-water partition coefficient ($P_\text{o/w}$) is defined as the equilibrium concentration ratio of a neutral solute between the solvents 1-octanol and water.
It is a signature of the relative affinity of a solute for an organic medium compared to an aqueous phase, as 1-octanol can represent soils, living tissues, and microorganisms \cite{Garrido_2009}.
Thus, $P_\text{o/w}$ has been used mainly to study systems of interest to the pharmaceutical industry, as in evaluating drug transfer between biological media \cite{Lipinski_2001, Perlovich_2004}.
Its experimental measurement has well-defined protocols, making the results very reliable \cite{Liao_2006}.

In the common logarithmic representation, the octanol-water partition coefficient is related to the hydration free energy of the solute ($\Delta G_\text{hyd}$) and its solvation free energy in 1-octanol ($\Delta G_\text{oct}$), both at temperature $T$ and atmospheric pressure, by \cite{Garrido_2009, Bannan_2016}
\begin{equation}
    \log{P_\text{o/w}}
    = \frac{\Delta G_\text{hyd} - \Delta G_\text{oct}}{k_\textsc{b} T \ln{10}}
    \label{eq:partition_coefficient}
\end{equation}
\citet{Bannan_2016} organized a database of simulation results for 36 small organic molecules as well as a comparison with experimental data found in the literature \cite{Hansch_1995, Sangster_1989, Leo_1971}.
This database was used as a reference for the studies carried out here.
Like them, we used the hypothesis of complete immiscibility between the solvents, although experimental data indicate non-negligible miscibility between water and 1-octanol \cite{Lang_2012}.
For each solvent, we consider a single solute molecule in the simulation box with periodic boundary conditions.
This is reasonable assuming that the experimental data to be compared were obtained at low solute concentrations.

\subsection{Simulation Details}
\label{sec:Simulation Details}

In the present study, we employed the highly efficient and customizable OpenMM package (version 7.5.0) \cite{Eastman_2017} for all of the MD simulations.
Some details of our OpenMM implementation are available in the Supporting Information.
The optimizations were performed with three different solute molecules: phenol, ibuprofen, and cyanuric acid.
Initial configurations, interaction parameters, and atomic charges
were taken from the FreeSolv database \cite{Mobley_2014}, which also provides experimental and calculated hydration free energy values.
These systems were modeled with the general Amber force field (GAFF) \cite{Wang_2004} parameters and AM1-BCC \cite{Jakalian_2000, *Jakalian_2002} charges for the solutes, as well as TIP3P \cite{Jorgensen_1983} parameters and charges for water.

We used the proposed LBF model with different switching functions and optimal parameters as well as the soft-core model implemented in ref~\citenum{Abreu_2020a} to calculate octanol-water partition coefficients.
In this case, we built initial configurations via PACKMOL \cite{Martinez_2009} and determined interaction parameters via LEaP \cite{Case_2021} by using the GAFF \cite{Wang_2004} and TIP3P \cite{Jorgensen_1983} force fields again.
With Antechamber \cite{Case_2021}, we applied the AM1-BCC method \cite{Jakalian_2000, *Jakalian_2002} to determine atomic charges for 1-octanol and the solutes vanillin, pentachlorophenol, nicotine, and trimethylamine.

In all alchemical protocols involving a single set of simulations to calculate capped, residual, and electrostatic potentials, we simulated states determined by 21 evenly spaced values of $\lambda$ from $0$ to $1$ ($\Delta \lambda = 0.05$).
In the protocols entailing separate simulations, we used five evenly spaced values for the residual and electrostatic potentials ($\Delta \lambda = 0.2$), while for the capped and soft-core potentials, we used 15 unevenly spaced values ($\Delta \lambda = 0.1$ for $0.4\leq\lambda\leq 0.9$ and $\Delta \lambda = 0.05$ otherwise).

The simulations were executed under the isothermal-isobaric (NPT) ensemble at $298.15$~K and $1$~atm.
Pressure control was done by a Monte Carlo barostat \cite{Aqvist_2004} with attempts to change the periodic box volume at every 50 fs, and the maximum allowable volume change was adjusted throughout the simulation to obtain an acceptance rate of around $50\%$. 
Temperature control was done via Langevin dynamics with a leapfrog, middle-type integration scheme (LFMiddle) \cite{Zhang_2019} and with a friction coefficient of $10$ ps$^{-1}$.
Each simulation consisted of $4$ ns of equilibration followed by $20$ ns of production for an aqueous medium or $40$ ns for a 1-octanol medium.
A $2$ fs time step was employed with samples collected every $200$ fs.
The nonbonded cutoff distance was $1.2$~nm, and a switching distance of $1.1$~nm was employed.
The simulation boxes had initial edge lengths between $3.5$~nm and $4.1$~nm.
Furthermore, we constrained the length of every chemical bond involving a hydrogen atom and considered rigid water molecules in accordance with the TIP3P model.

When the PME method was used for the long-range electrostatic interactions, the Ewald damping parameter was assigned as $\alpha = \sqrt{-\ln{(2 \delta)}}/r_c$, with an error tolerance $\delta = 5 \times 10^{-4}$.
Moreover, the number of nodes in the mesh was determined as $n_\text{mesh} = (2/3)\alpha L_{\rm box} \delta^{-1/5}$, where $L_{\rm box}$ is the length of the box along each dimension.

Coupling free energies and average properties were estimated using the MBAR method as implemented in the PYMBAR library \cite{Shirts_2008, Chodera_2007}.
For decorrelation of the sampled properties, we used the overlapping batch-mean (OBM) estimator \cite{Meketon_1994} with blocks of size $b = \lfloor \sqrt{n} \rfloor$, where $n$ is the original sample size \cite{Flegal_2010}.
It should be noted that we used this method instead of the integrated autocorrelation function method implemented in PYMBAR \cite{Chodera_2007}. 
Uncertainties in free energies were computed using the MBAR estimator \cite{Shirts_2008, Kong_2003}, their propagation was handled via the delta method \cite{Greene_2012}, and their values are reported in the form of 95\% confidence intervals.

\section{Results and Discussion}
\label{sec:results}

\subsection{Modeling of Solute-Solvent Electrostatic Interactions}
\label{sec:solute-solvent electrostatic interactions}

Table~\ref{tab:PME_CRF_results} shows results for solutes of diverse chemical classes, 
whose simulated hydration free energies reported in the FreeSolv database \cite{Mobley_2014} ($\Delta G_{\rm hyd}^\textsc{fs}$) span a wide range, from $-9$ to $-93$~kJ/mol, as one can see in the second column of the table.

As described in section~\ref{sec:electrostatic}, for each solute we carried out separate simulations with the electrostatic solute-solvent interaction forces derived from either the CRF potential ($U_\textsc{crf}^\textsc{as}$) or the PME potential ($U_\textsc{pme}^\textsc{as}$).
The former simulations were computationally cheaper than the latter.
During these simulations, the values of both $U_\textsc{crf}^\textsc{as}$ and $U_\textsc{pme}^\textsc{as}$ were sampled at regular intervals.
Then, in a postprocessing stage, the free energy difference $G^\textsc{as}_\textsc{pme}-G^\textsc{as}_\textsc{crf}$ was estimated using both the BAR and FEP methods.
This difference is the correction required for estimating standard PME-based hydration free energies from values obtained using the CRF model to evaluate solute-solvent interactions instead.
The results are shown in the two rightmost columns of Table~\ref{tab:PME_CRF_results}.

All of the studied substances whose simulated hydration-free energies lie above $-46$~kJ/mol require minimal CRF-to-PME corrections ($< 0.4$~kJ/mol).
It is important to remark that more than 95\% of the compounds in the current FreeSolv database (611 out of 642) have $\Delta G_{\rm hyd}^\textsc{fs} > - 46$~kJ/mol.
For cyanuric acid, the compound with the most negative value of $\Delta G_{\rm hyd}^\textsc{fs}$, the required correction is only $1.1$~kJ/mol.
Moreover, the correction itself can be accurately obtained from unilateral FEP calculations since the values computed in this way are in
agreement with those from the bilateral BAR method.
Therefore, we only need to run the cheaper simulation in which the CRF forces are computed at every step but $U_\textsc{crf}^\textsc{as}$ and $U_\textsc{pme}^\textsc{as}$ are computed only from time to time for sampling.
This analysis shows that the reaction field model with infinite dielectric constant can satisfactorily substitute PME calculations when we need to compute solute-solvent interactions separately.

\begin{table*}[htb!]
\footnotesize
	\centering
	\setlength\tabcolsep{0pt}
	\caption{Free energy corrections to be applied when different types of pairwise potentials replace particle mesh Ewald calculations in the evaluation of solute-solvent electrostatic interactions.$^a$ }
	\label{tab:PME_CRF_results}
	\begin{tabular*}{\textwidth}{
		@{\extracolsep{\fill}}
		l
		S[table-format=-2.2(0)]
		S[table-format=-2.3(3)]
		S[table-format=-2.3(3)]
		S[table-format=-2.3(3)]
		S[table-format=-2.3(3)]
		S[table-format=-2.3(3)]
	}
		\hline
		Solute &
		{$\Delta G_{\rm hyd}^\textsc{fs}$} &
		{$G^\textsc{as}_{\textsc{pme}}-G^\textsc{as}_{\textsc{dsf}}$} &
		{$G^\textsc{as}_\textsc{pme}-G^\textsc{as}_\textsc{sf}$} &
		{$G^\textsc{as}_\textsc{pme}-G^\textsc{as}_\textsc{rf}$} &
		{$G^\textsc{as}_\textsc{pme}-G^\textsc{as}_\textsc{crf}$} &
		{$[G^\textsc{as}_\textsc{pme}-G^\textsc{as}_\textsc{crf}]_\textsc{fep}$} \\
		\hline
		urea          & -9.00  & -2.707\pm0.007 & -0.193\pm0.003 & -1.484\pm0.004 & -0.216\pm0.003 & -0.224\pm0.005 \\
		formaldehyde  & -13.18 & -2.068\pm0.004 &  0.071\pm0.004 & -1.276\pm0.003 &  0.024\pm0.003 &  0.010\pm0.005 \\
		phenol        & -23.89 & -2.477\pm0.004 &  0.162\pm0.004 & -1.725\pm0.003 &  0.116\pm0.004 &  0.106\pm0.006 \\
		DES           & -36.32 &  -4.18\pm0.04  &  0.533\pm0.009 &  -2.50\pm0.02  &  0.391\pm0.008 &   0.39\pm0.02  \\
		glycerol      & -42.43 &  -3.13\pm0.02  &  0.296\pm0.007 &  -2.75\pm0.01  &  0.183\pm0.006 &   0.18\pm0.01  \\
		MSM           & -44.18 & -4.979\pm0.008 &   0.62\pm0.01  & -2.715\pm0.007 &  0.392\pm0.009 &   0.35\pm0.02  \\
		ibuprofen     & -45.44 &  -6.82\pm0.04  & -0.220\pm0.007 & -4.479\pm0.009 & -0.337\pm0.007 &  -0.36\pm0.01  \\
		AAQ           & -58.12 & -5.692\pm0.008 &  0.959\pm0.009 & -2.986\pm0.006 &  0.788\pm0.008 &   0.76\pm0.02  \\
		caffeine      & -73.72 & -3.773\pm0.006 &   0.90\pm0.01  & -2.996\pm0.007 &  0.628\pm0.009 &   0.59\pm0.02  \\
		cyanuric acid & -93.04 & -1.184\pm0.004 &  1.441\pm0.009 & -2.290\pm0.006 &  1.111\pm0.008 &   1.08\pm0.02  \\
		\hline
	\end{tabular*}
{\footnotesize $^a$ The chemical species were carefully selected from the FreeSolv database. Values are reported in  kJ/mol. Free energy differences were computed using the BAR (columns 3-6) and FEP (column 7) methods. Abbreviations: DES=diethyl succinate; MSM=methylsulfonylmethane; AAQ=2-aminoanthraquinone}
\end{table*}

We also compared the CRF model with other pairwise alternatives.
Eq~\eqref{eq:CRF_reduce} is the $\epsilon_\textsc{rf}\to\infty$ limit of a more general reaction field (RF) model, $\nu_\textsc{rf}(x) = 1/x + [( \epsilon_\textsc{rf}-1)x^2-3\epsilon_\textsc{rf}]/(2\epsilon_\textsc{rf} + 1)$, where $\epsilon_\textsc{rf}$ is the dielectric constant of the medium outside the cutoff cavity \cite{Tironi_1995}.
In this model, a dielectric constant equal to 1 (vacuum) is considered inside the cavity.
We tested it with $\epsilon_\textsc{rf}=78.3$, which is the dielectric constant of water at room temperature.
Although the derivative $\nu'_\textsc{rf}(x)$ fails to vanish at $x=1$ when $\epsilon_\textsc{rf}$ is finite \cite{Kubincova_2020}, this is not a problem here due to the presence of a switching function in eq~\eqref{eq:crf_complete}.
We also tested the shifted force (SF) version of the Coulomb potential, which can successfully replace lattice-sum methods in numerous situations \cite{Fennell_2006a}.
It consists in replacing $\nu_\textsc{crf}(x)$ in eq~\eqref{eq:crf_complete} by $\nu_\textsc{sf}(x) = 1/x + x - 2$, which satisfies $\nu_\textsc{sf}(1) = \nu_\textsc{sf}'(1) = 0$ as well.
In fact, this is a special case of the damped shifted force (DSF) model \cite{Fennell_2006a},
\begin{multline*}
\nu_\textsc{dsf}(x) =  \frac{\text{erfc}(\alpha r_c x)}{x} - \text{erfc}(\alpha r_c) + \\
(x-1) \left[\text{erfc}(\alpha r_c) + \frac{2 \alpha r_c}{\sqrt{\pi}} \text{erfc}(-\alpha^2 r_c^2) \right]
\end{multline*}
which we also tested here with a damping parameter $\alpha = 0.2$ {\AA}$^{-1}$.

\begin{figure*}[htb!]
	\centering
	\includegraphics[scale=1.0]{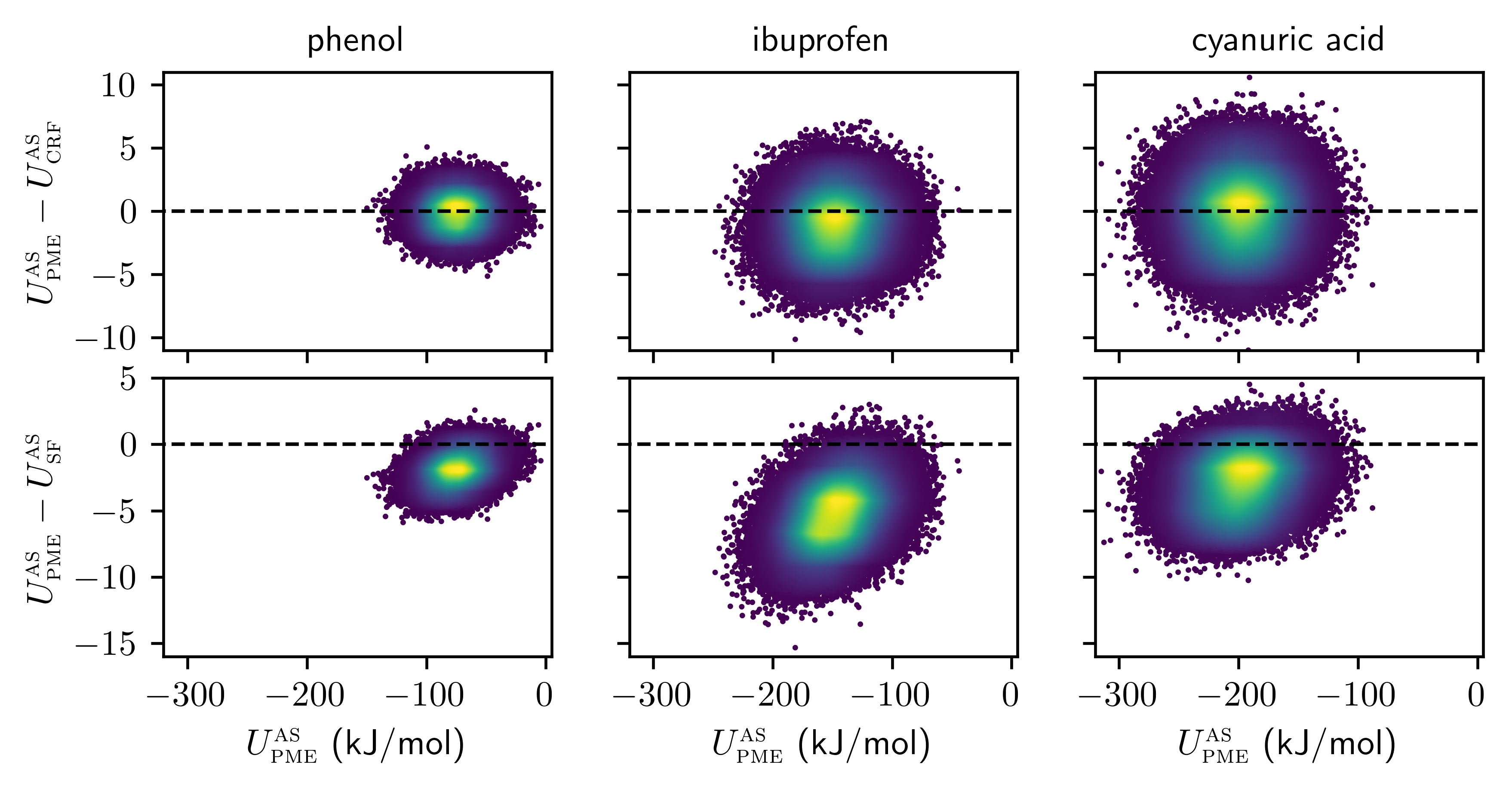}
	\caption{
		Comparison between solute-solvent electrostatic potential energies computed for the same configurations by the PME method and a pairwise alternative: the conductor-like reaction field (CRF) model (top panel) or the shifted force (SF) potential (bottom panel).
		The configurations were sampled during MD simulations based on the PME method to compute forces.
		The warmer the color tone corresponding to a potential pair, the greater is the probability of finding that pair during the simulation.
		}
	\label{fig:reac_fields}
\end{figure*}

In columns 3, 4, and 5 of Table~\ref{tab:PME_CRF_results}, we can see BAR-estimated values of the corrections needed to obtain PME-based solvation free energies from simulations applying the DSF, SF, and RF models, respectively, for computing the solute-solvent interactions.
Only the SF model exhibits performance comparable to that of the CRF model.
The needed corrections for the remaining ones are systematically much larger, making them unsuitable for this type of calculation.
Table~\ref{tab:PME_CRF_results} suggests that the CRF model is still slightly better than the SF model, but a more detailed comparison is needed to attest to its superiority.
Our analysis consists in plotting coordinate pairs $(U_\textsc{pme}^\textsc{as}, U_\textsc{pme}^\textsc{as}-U_\textsc{crf}^\textsc{as})$ and $(U_\textsc{pme}^\textsc{as}, U_\textsc{pme}^\textsc{as}-U_\textsc{sf}^\textsc{as})$ sampled during a simulation employing forces derived from $U_\textsc{pme}^\textsc{as}$.
If these pairwise models are suitable substitutes for PME calculations, those coordinate pairs are expected to lie close to the horizontal lines $U_\textsc{pme}^\textsc{as} - U_\textsc{crf}^\textsc{as} = 0$ and $U_\textsc{pme}^\textsc{as} - U_\textsc{sf}^\textsc{as} = 0$, respectively.
These plots are shown in Figure~\ref{fig:reac_fields} for three selected solutes covering a wide range of solvation free energies, namely, phenol, ibuprofen, and cyanuric acid.
The color of each plotted coordinate pair is picked from a dark-to-light palette according to the value of a kernel density estimate (KDE) of the sample probability distribution.
To obtain the KDE, we employed Gaussian kernels with bandwidths determined via Scott's rule \cite{Scott_1992}.
In all cases, the CRF model exhibits a high probability that $U_\textsc{pme}^\textsc{as} - U_\textsc{crf}^\textsc{as} \approx 0$ with a symmetric distribution around $U_\textsc{crf}^\textsc{as} = U_\textsc{pme}^\textsc{as}$.
For the SF model, however, we observe a marked tendency of $U_\textsc{sf}^\textsc{as}$ to overestimate the value of $U_\textsc{pme}^\textsc{as}$, which makes it a less reliable substitute than the CRF model.

We used the CRF model and the subsequent correction to the PME model in all of the simulations involving solvation free energy calculations in this work, even for 1-octanol solvent instead of water. This includes the simulations with both the LBF and the soft-core approaches to validate our new proposal.

\subsection{Optimization Analysis}
\label{sec:optimization_results}

This section presents and discusses the results of the optimization procedures described in section~\ref{sec:optimizations}.
They are applied independently for the solutes phenol, ibuprofen, and cyanuric acid in an aqueous phase.
Nonetheless, we intend to find sets of switching function parameters that perform well in all three cases.
We expect these consensus parameters to provide reasonable alchemical pathways for a wide range of systems besides the analyzed ones.
As shown in section~\ref{sec:free_energy_difference_results}, the same parameters are suitable for calculations having 1-octanol as the solvent.

\subsubsection{Smooth Switching Functions}

\begin{figure*}[htb]
	\centering
	\includegraphics[scale=1.0]{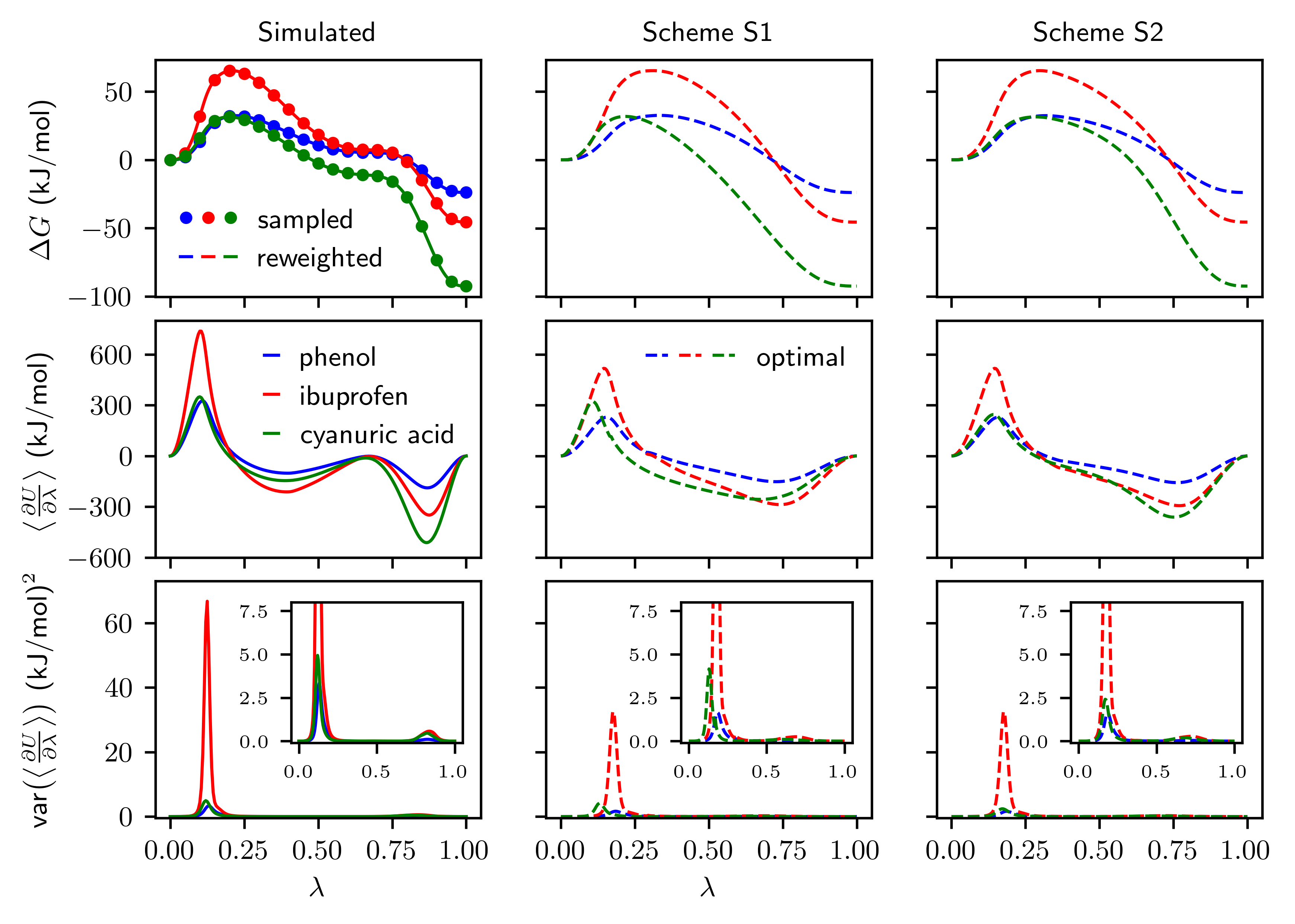}
	\caption{
    Property profiles for the hydration of multiple solutes obtained by a concerted pathway with smooth switching functions.
    Circles and solid lines indicate simulation and reweighting results, respectively, for the reference system.
    The dashed lines represent reweighting results at states defined by the optimal switching functions obtained through schemes S1 and S2.
	}
	\label{fig:op1and2}
\end{figure*}

We present the optimization results obtained with schemes S1 and S2 described in section~\ref{sec:optimizations}.
Both depend on the same MD simulations performed with the reference switching functions whose parameters were set as the initial guess for each scheme.
In the first column of Figure~\ref{fig:op1and2}, one can see the coupling free energy profiles obtained in these simulations as well as the mean derivatives $\langle \partial U/\partial \lambda \rangle$ and free energy variances as functions of $\lambda$.
For all three solutes, virtually constant free energies are observed between $\lambda = 0.5$ and $\lambda = 0.75$, creating a local maximum close to zero in the $\langle \partial U/\partial \lambda \rangle$ curve.
In contrast, a sharp decay in $\Delta G$, followed by a slower decay and final stabilization, occurs at $0.75 < \lambda \leq 1$, leading to a deep well in the $\langle \partial U/\partial \lambda \rangle$ curve.
This behavior indicates that the reference alchemical pathway is inefficient because too many intermediate states are allocated to a $\lambda$ range with little variation in free energy and too few states are allocated to a range with high variation.
To overcome this issue, minimizing sudden changes in the derivative $\langle \partial U/\partial \lambda \rangle$, as done in scheme S1, has been a suitable approach.

The dashed lines in the second column of Figure~\ref{fig:op1and2} result from the optimal switching functions obtained by applying scheme S1.
As done in every optimization step, including the one responsible for finding the optimal solution, these curves were inferred via reweighting.
As desired, they show a more uniform free energy profile and a $\langle \partial U/\partial \lambda \rangle$ curve without unneeded local maxima and narrow wells.
We checked the effectiveness of reweighting by performing new simulations with the optimal parameters.
Their results, which we have omitted to avoid redundancy, perfectly agreed with the reweighted curves.

Returning to the first column of Figure~\ref{fig:op1and2}, we now analyze the free energy variance for the reference system.
Considerably higher variance values occur between $\lambda = 0$ and $\lambda = 0.25$, where a free energy barrier is present.
Thus, another way to improve the alchemical pathway is by reducing the variance in this region.
We expect this to occur when the area under the variance curve is minimized in scheme S2.
The third column of Figure~\ref{fig:op1and2} contains curves obtained by reweighting to a state defined by the optimal switching functions found with scheme S2.
Again, we validated these curves through new simulations.
The free energy barrier between $\lambda = 0$ and $\lambda = 0.25$ exhibits a more gradual variation, and the area under the variance curve is considerably diminished.
The gentler rise of the free energy curve toward its maximum relates to the shrinking of the hump in the mean derivative curve, whose peak is aligned to that in the variance curve.

The optimization involving a space with fewer dimensions was preferable for both schemes because of the similarity of its results compared to the case with more parameters.
Thus, all analyses in this section refer to the optimizations in which $\lambda_\textsc{r}^\text{s} = \lambda_\textsc{e}^\text{s}$.
The optimal parameters are presented in Table~\ref{tab:obj_func_op1_op2}, in which $f_{\rm obj}^{\rm init}$ and  $f_{\rm obj}^{\rm opt}$ refer to objective function values for the reference system and the optimal parameters, respectively.

\begin{table}[htb!]
	\centering
	\setlength\tabcolsep{0pt}
	\caption{
		Optimal parameter values and initial and minimum objective function values obtained using schemes S1 and S2 for the concerted alchemical pathway with smooth switching functions.$^a$
	}
	\label{tab:obj_func_op1_op2}
	\begin{tabular*}{\linewidth}{@{\extracolsep{\fill}}llcccc}
		\hline
		Scheme & Solute  & $\lambda_\textsc{c}^\text{f}$ & $\lambda_\textsc{r}^\text{s} = \lambda_\textsc{e}^\text{s}$ & $f_{\rm obj}^{\rm init}$ & $f_{\rm obj}^{\rm opt}$ \\
		\hline
		\multirow{3}{*}{S1} & Phenol & 0.994 & 0.293 & 1229 & 763  \\
		 & Ibuprofen & 1.000 & 0.294 & 2593 & 1610 \\
		 & Cyanuric acid & 0.762 & 0.165 & 1991 & 1155 \\  
		\hline
		\multirow{3}{*}{S2} & Phenol & 1.000 & 0.366 & 0.175 & 0.122   \\
		& Ibuprofen & 1.000 & 0.359 & 1.745 & 1.226 \\
		& Cyanuric acid & 1.000 & 0.329 & 0.261 & 0.183 \\ 
		\hline
	\end{tabular*}
	{\footnotesize $^a$ The objective functions of these optimization schemes are measured in kJ/mol and (kJ/mol)\textsuperscript{2}, respectively.}
\end{table}

\begin{figure*}[htb]
	\centering
	\includegraphics[scale=1.0]{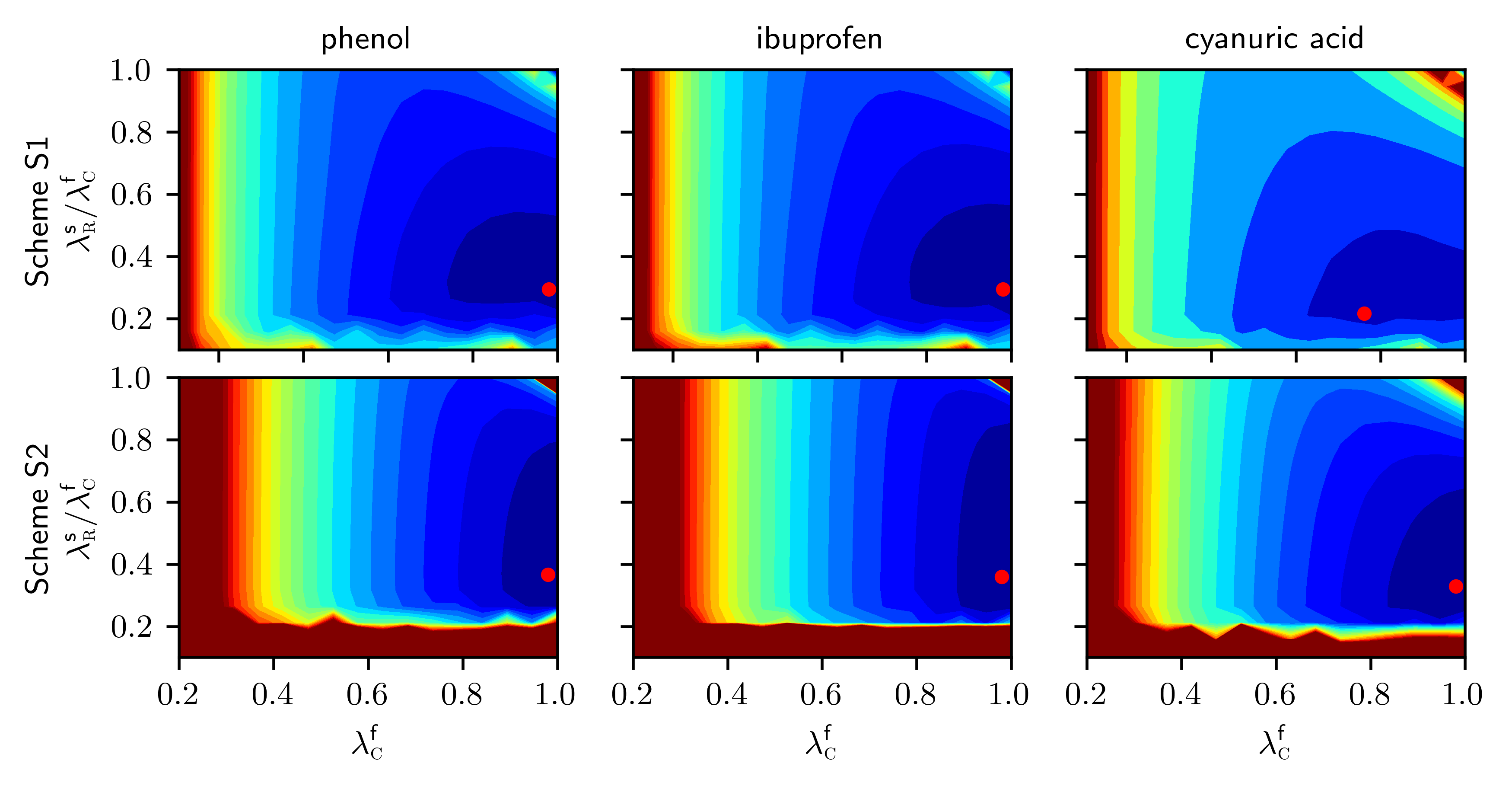}
	\caption{
	    Objective function contour plots for different solutes in water.
		The first and second rows were obtained using optimization schemes S1 and S2, respectively.
		Dark-blue regions contain objective function values smaller than 10\% above the minimum.
		Red dots correspond to the optimal parameter values.
		}
	\label{fig:op_curvas}
\end{figure*}

In Figure~\ref{fig:op_curvas} we present contour plots of the objective functions of both schemes in the space $\lambda_\textsc{r}^\text{s}/\lambda_\textsc{c}^\text{f}$ versus $\lambda_\textsc{c}^\text{f}$.
The optimal parameters values in Table~\ref{tab:obj_func_op1_op2} are indicated as red dots.
The darker-blue region contains the points considered to be nearly optimal, whose $f_{\rm obj}$ values are not larger than 10\% above the minimum.
This region has a considerable extension for all cases, leading to the possibility of a wide family of effective alchemical pathways.

The optimal pairs $(\lambda_\textsc{c}^\text{f}, \lambda_\textsc{r}^\text{s})$ obtained using scheme S1 for phenol and ibuprofen were close to each other but considerably distant from that obtained for cyanuric acid.
On the other hand, scheme S2 led to very similar optimal pairs for all three solutes.
It should be noticed that the geometric center of these pairs lies at $(1.0, 0.35)$, which is located in the dark-
blue region in all six plots of Figure~\ref{fig:op_curvas}.
For this reason, we chose the values $\lambda_\textsc{c}^\text{f} = 1.0$ and $\lambda_\textsc{r}^\text{s} = \lambda_\textsc{e}^\text{s} = 0.35$ as the consensus parameters for the smooth switching functions used in our concerted LBF formulation.

We expect the consensus parameters to provide effective alchemical pathways for solvation free energy calculations with diverse solutes and solvents.
If this is not true for a specific case, using one of the optimization procedures discussed here can provide parameters for the desired system.
Although the consensus parameters ultimately stemmed from scheme S2, it is worth noting that the more straightforward objective function of scheme S1 was also effective in finding efficient parameter sets.

\subsubsection{Quartic Switching Function}

\begin{figure}[htb!]
	\centering
	\includegraphics[scale=1.0]{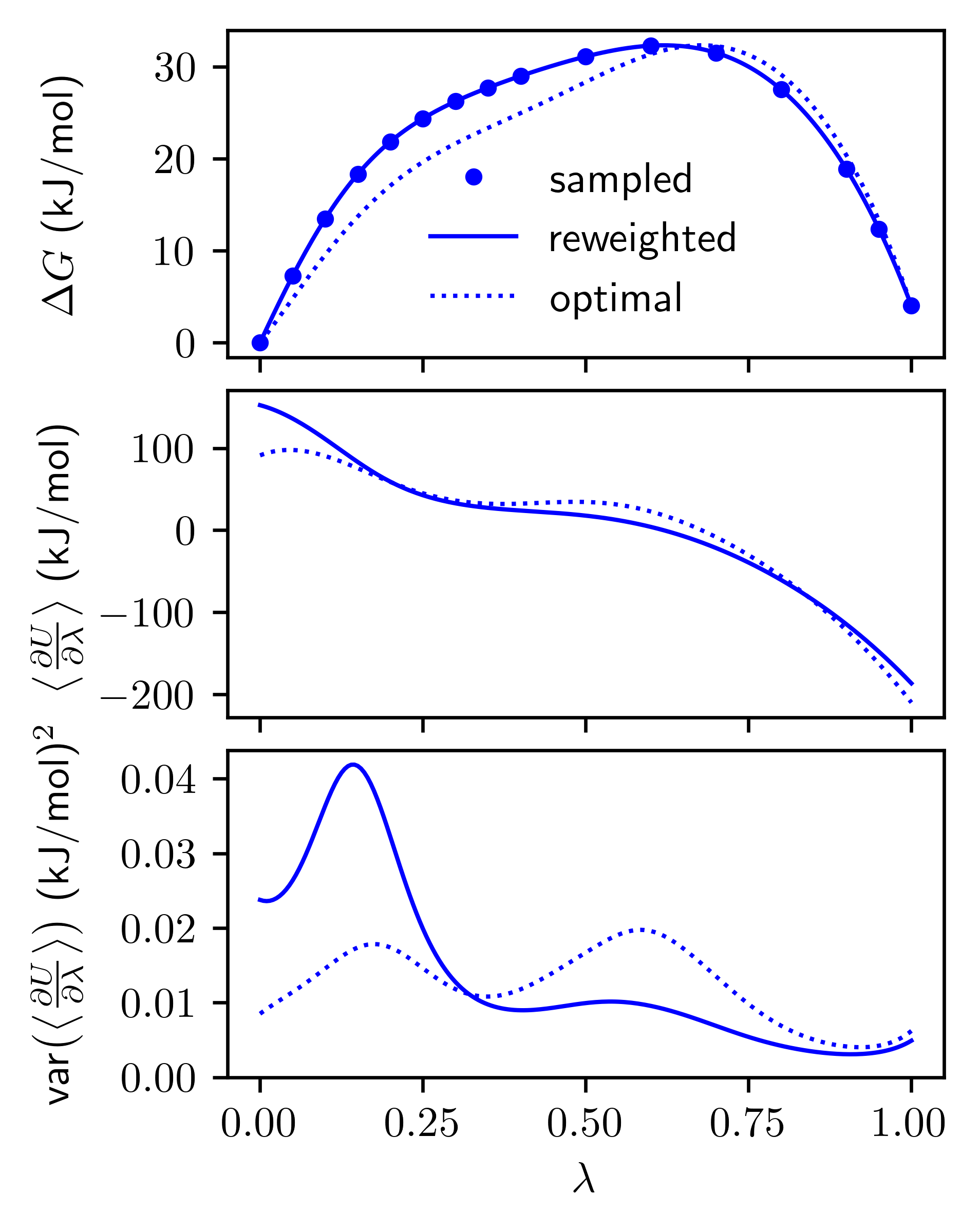}
	\caption{
Property profiles regarding the coupling of the capped LJ potential with a quartic switching function in the hydration of phenol.
Circles and solid lines indicate simulation and reweighting results, respectively, for the reference system.
Dashed lines represent reweighting results in the case of the optimal switching function derived from scheme Q.	}
	\label{fig:cap}
\end{figure}

We performed independent simulations for the capped, residual, and electrostatic alchemical potentials to apply the scheme Q optimization procedure, which involves the quartic switching function.
Figure~\ref{fig:cap} presents results for the capped potential for phenol.
We used this solute as an example, but the others behave similarly, as shown in analogous figures made available in the Supporting Information.
In the region between $\lambda = 0$ and $\lambda=0.25$, the free energy increases more gradually with the optimal switching functions than with those of the reference system.
In the region where $\lambda > 0.3$, we observe an increase in the local variance, which was not as significant as the reduction in the preceding region.

\begin{figure}[htb!]
	\centering
	\includegraphics[scale=1.0]{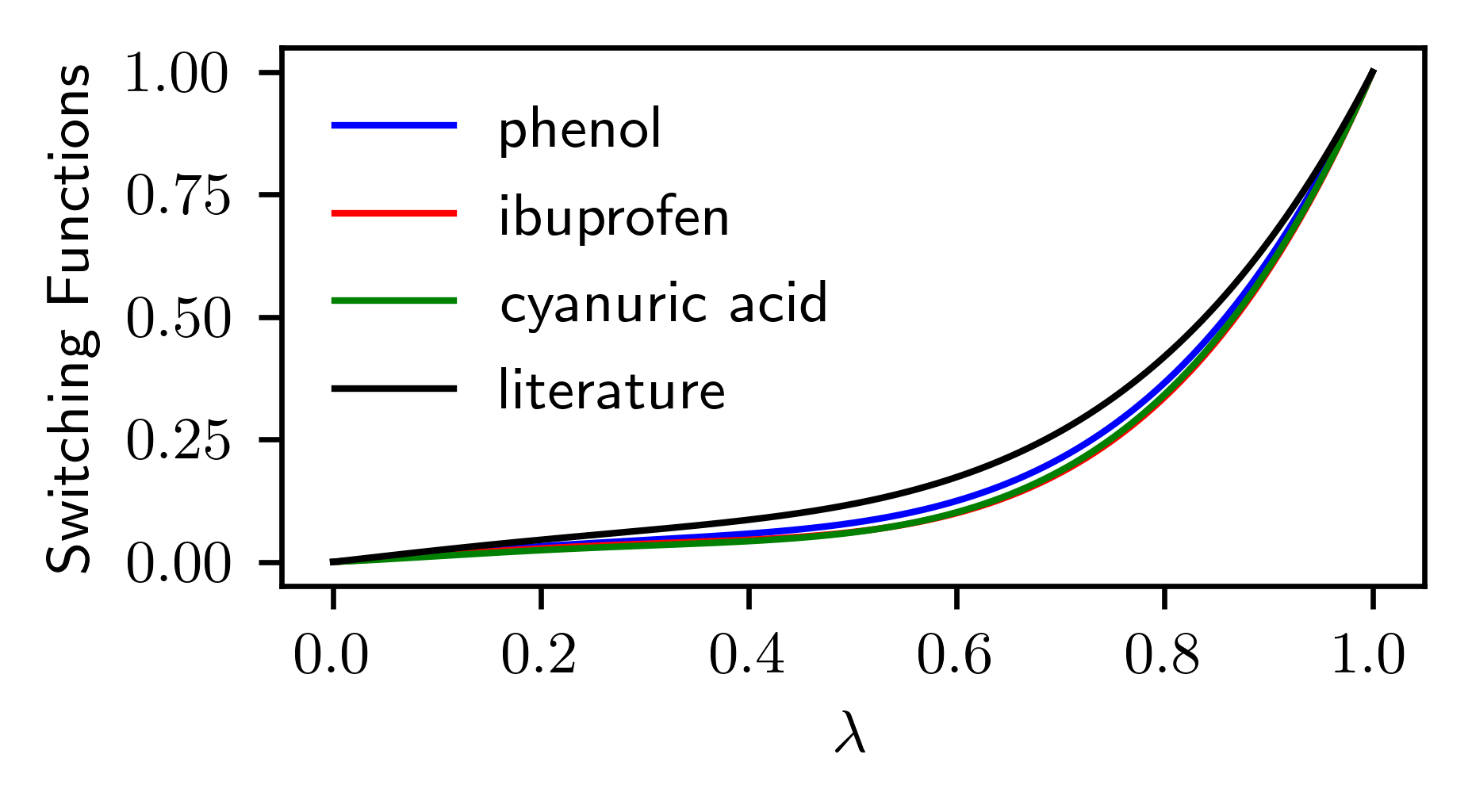}
	\caption{Optimal switching functions obtained for the capped potential in the sequential LBF approach and comparison with the one used in ref~\citenum{Naden_2014}.}
	\label{fig:switching}
\end{figure}

Table~\ref{tab:schemeIIItable} shows the optimal parameters and the variance values before ($f_\text{obj}^\text{init}$) and after ($f_{\rm obj}^{\rm opt}$) optimization.
In general, the residual and electrostatic potentials have a small influence on the global variance of the alchemical pathway.
Thus, following Naden et al. \cite{Naden_2014, Naden_2015}, we recommend the use of linear switching functions in these cases for the sake of simplicity.
For the capped potential, we will adopt as a consensus the optimal parameter values obtained for ibuprofen, $A_\textsc{c} = 2.568$, $B_\textsc{c} = -2.104$, and $C_\textsc{c} = 0.419$, which show good agreement with those of the other solutes, as noted in Figure~\ref{fig:switching}.
When contrasted to phenol and cyanuric acid, ibuprofen is a larger molecule with an intermediate hydration free energy value.

\begin{table*}[htb!]
	\centering
	\setlength\tabcolsep{0pt}
	\caption{
		Optimal parameters and initial and minimum objective function values obtained using scheme Q for the sequential alchemical pathway with quartic switching functions.$^a$
	}
	\label{tab:schemeIIItable}
	\begin{tabular*}{\textwidth}{
			@{\extracolsep{\fill}}
			llccccc}
		\hline
		
		Potential & 
		Solute &                                 
		$A_k$ &
		$B_k$ &
		$C_k$ &
		$f_{\rm obj}^{\rm init} \times 10^3$ &
		$f_{\rm obj}^{\rm opt} \times 10^3$ \\
		
		\hline
		
		\multirow{3}{*}{Capped} & Phenol & \num{2.232} & \num{-1.694} & \num{0.316} & \num{13.8} & \num{12.4} \\
		& Ibuprofen & \num{2.568} & \num{-2.104} & \num{0.419} & \num{74.4} & \num{54.3} \\
		& Cyanuric acid & \num{2.474} & \num{-2.010} & \num{0.445} & \num{19.7} & \num{14.1}  \\

		\hline

		\multirow{3}{*}{Residual} & Phenol & \num{0.132} & \num{0.247} & \num{0.482} & \num{0.0552} & \num{0.0314}  \\
		& Ibuprofen & \num{-0.0595} & \num{0.319} & \num{0.567} & \num{0.184} & \num{0.120} \\
		& Cyanuric acid & \num{0.00923} & \num{0.228} & \num{0.437} & \num{0.0143} & \num{0.0112} \\

		\hline

		\multirow{3}{*}{Electrostatic} & Phenol & \num{-0.182} & \num{1.014} & \num{-1.874} & \num{3.93} & \num{3.39}  \\
		& Ibuprofen & \num{-0.310} & \num{1.854} & \num{-3.196} & \num{28.4} & \num{20.9} \\
		& Cyanuric acid & \num{-0.0712} & \num{0.827} & \num{-1.809} & \num{16.1} & \num{13.7} \\

		\hline
	\end{tabular*}
	{\footnotesize $^a$ The objective function of this optimization scheme is measured in (kJ/mol)\textsuperscript{2}.}
\end{table*}

\subsection{Octanol-Water Partition Coefficient}
\label{sec:free_energy_difference_results}

We used eq~\eqref{eq:partition_coefficient} to calculate octanol-water partition coefficients from solvation free energies determined in both water and 1-octanol media for the solutes trimethylamine (TMA), nicotine, pentachlorophenol (PCP), and vanillin.
We obtained such free energy values through two different alchemical pathways, formulated with the switching functions and their consensus parameters discussed in section~\ref{sec:optimization_results}.
We also used the ubiquitous soft-core model for the LJ potential to validate the newly proposed methodology.

Table~\ref{tab:energy_total} shows the solvation free energies in water and 1-octanol for the four analyzed solutes and each different alchemical pathway.
We can see a good agreement among the alchemical pathways, with uncertainties of similar magnitudes.
The table also displays simulation results reported by \citet{Bannan_2016}.
The observed deviations are probably due to differences in details of the adopted MD procedures, mainly in the atomic charges used for the solutes and 1-octanol.
We note that the authors deliberately used OpenEye's QUACPAC application \cite{OEChem} instead of Antechamber \cite{Case_2021} to apply the AM1-BCC method.
Since the partial charges of the water model do not come from this method, the influence of charge dissimilarity would be better quantified if we could compare the LJ contributions to the hydration free energies.
However, this specific information is not available in ref~\citenum{Bannan_2016}.

We remark that although the optimal quartic switching functions yield partial coupling free energies with low variances, as shown in Table~\ref{tab:schemeIIItable}, error propagation leads to overall free energy uncertainties that are only slightly smaller than those resulting from the optimal concerted coupling.
Such an achievement of schemes S1 and S2 is outstanding, considering that it involves only two optimization parameters while scheme Q involves nine parameters for a full solute-solvent coupling.
The Supporting Information contains two figures that depict, for the aqueous and 1-octanol systems, how the uncertainties vary with simulated time for each method analyzed in Table~\ref{tab:energy_total}.
They show that the concerted LBF method is on par with the others in terms of computational efficiency.

\begin{table*}[htb!]
	\centering
	\setlength\tabcolsep{0pt}
	\caption{Solvation free energy values (in kJ/mol) in water and 1-octanol for different solutes obtained using the soft-core
model and the sequential and concerted LBF alchemical pathways; data found in the literature are also reported.}
	\label{tab:energy_total}
	\begin{tabular*}{\textwidth}{@{\extracolsep{\fill}}llcccc}
		\hline
		Solute & Solvent & Soft-core & Sequential LBF & Concerted LBF & ref~\citenum{Bannan_2016} \\ 		
		\hline
		\multirow{2}{*}{TMA} & Water & $-11.4 \pm 0.1$  & $-11.2 \pm 0.1$ & $-11.2 \pm 0.2$ & $-11.2 \pm 0.1$\\
		& 1-Octanol & $-16.3 \pm 0.3$  & $-16.7 \pm 0.3$ & $-16.6 \pm 0.2$ & $-17.4 \pm 0.2$ \\
		\hline
		\multirow{2}{*}{Nicotine} & Water & $-29.1 \pm 0.2$  & $-29.0 \pm 0.2$ & $-29.4 \pm 0.3$ & $-29.2 \pm 0.1$ \\
		& 1-Octanol & $-43.4 \pm 0.4$  & $-43.2 \pm 0.5$ & $-43.7 \pm 0.6$ & $-47.5 \pm 0.3$ \\
		\hline
		\multirow{2}{*}{PCP} & Water & $-13.8 \pm 0.2$  & $-13.8 \pm 0.2$ & $-13.8 \pm 0.3$ & $-14.6 \pm 0.1$ \\
		& 1-Octanol & $-51.6 \pm 0.4$  & $-51.8 \pm 0.4$ & $-52.0 \pm 0.5$ & $-51.9 \pm 0.3$ \\
		\hline
		\multirow{2}{*}{Vanillin} & Water & $-35.8 \pm 0.2$  & $-35.8 \pm 0.2$ & $-35.5 \pm 0.2$ & $-33.8 \pm 0.1$\\ 
		& 1-Octanol & $-44.5 \pm 0.5$  & $-45.3 \pm 0.4$ & $-45.0 \pm 0.4$ & $-43.5 \pm 0.3$ \\ 
		\hline& 
	\end{tabular*}
\end{table*}

Table~\ref{tab:schemeSandQ} presents more details for comparing the concerted LBF approach using smooth switching functions to the sequential one using quartic functions.
It contains the capped ($\Delta G_\textsc{c}$), residual ($\Delta G_\textsc{r}$), and electrostatic ($\Delta G_\textsc{e}$) contributions to the solvation free energies.
In the case of the concerted approach, these contributions were computed by reweighting the simulation data to the milestone states of the sequential pathway, which are those defined by $(h_\textsc{c}, h_\textsc{r}, h_\textsc{e}) = (1,0,0)$ and $(1,1,0)$.
We can see a good agreement between the free energy components for the two LBF approaches. The uncertainty is systematically smaller in the sequential case for the residual term, which shows that the number of states used for this calculation was oversized.
The Supporting Information also discriminates the LJ and electrostatic contributions to the solvation free energies obtained from soft-core-based simulations. A single electrostatic contribution calculation applies for both the sequential and soft-core pathways, as a linear coupling was performed for these cases. The soft-core strategy refers only to the coupling of the LJ potential.

\begin{table*}[htb!]
	\centering
	\setlength\tabcolsep{0pt}
	\caption{Free energy values (in kJ/mol) for the coupling of capped, residual, and electrostatic potentials in the application of the sequential and concerted LBF alchemical pathways; the concerted values were calculated in terms of the milestone states of the sequential pathway.}
	\label{tab:schemeSandQ}
	\begin{tabular*}{\textwidth}{@{\extracolsep{\fill}}llcccccc}
		\hline
		\multirow{3}{*}{Solute} & \multirow{3}{*}{Pathway} & \multicolumn{3}{c}{Water as solvent} & \multicolumn{3}{c}{1-Octanol as solvent} \\
		\cline{3-5} \cline{6-8}
		&
		& $\Delta G_\textsc{c}$ 
		& $\Delta G_\textsc{r}$
		& $\Delta G_\textsc{e}$
		& $\Delta G_\textsc{c}$
		& $\Delta G_\textsc{r}$
		& $\Delta G_\textsc{e}$
		\\
		\hline
		\multirow{2}{*}{TMA}      & Sequential & $5.5 \pm 0.1$ & $1.68 \pm 0.01$ & $-18.5 \pm 0.1$ & $-13.7 \pm 0.2$ & $4.09 \pm 0.02$ &  $-7.1 \pm 0.3$ \\
		                          & Concerted  & $5.6 \pm 0.1$ &  $1.7 \pm 0.1$  & $-18.5 \pm 0.1$ & $-13.4 \pm 0.1$ &  $4.2 \pm 0.2$  &  $-7.3 \pm 0.2$ \\
		\hline
		\multirow{2}{*}{Nicotine} & Sequential & $1.0 \pm 0.2$ & $3.08 \pm 0.02$ & $-33.1 \pm 0.1$ & $-41.4 \pm 0.3$ & $8.19 \pm 0.04$ & $-10.0 \pm 0.4$ \\
		                          & Concerted  & $0.9 \pm 0.1$ &  $3.1 \pm 0.2$  & $-33.3 \pm 0.2$ & $-42.1 \pm 0.3$ &  $8.3 \pm 0.4$  &  $-9.9 \pm 0.4$ \\
		\hline
		\multirow{2}{*}{PCP}      & Sequential & $0.8 \pm 0.2$ & $0.74 \pm 0.01$ & $-15.4 \pm 0.1$ & $-44.9 \pm 0.3$ & $3.32 \pm 0.05$ & $-10.2 \pm 0.3$ \\
		 	                      & Concerted  & $0.8 \pm 0.1$ &  $0.7 \pm 0.2$  & $-15.4 \pm 0.2$ & $-45.0 \pm 0.2$ &  $3.3 \pm 0.3$  & $-10.3 \pm 0.3$ \\
		\hline
		\multirow{2}{*}{Vanillin} & Sequential & $1.4 \pm 0.2$ & $1.79 \pm 0.01$ & $-39.0 \pm 0.2$ & $-36.1 \pm 0.3$ & $5.15 \pm 0.03$ & $-14.3 \pm 0.4$ \\
		                          & Concerted  & $1.4 \pm 0.1$ &  $1.8 \pm 0.2$  & $-38.7 \pm 0.2$ & $-36.4 \pm 0.2$ &  $5.2 \pm 0.2$  & $-13.8 \pm 0.3$ \\
		\hline
	\end{tabular*}
\end{table*}

\begin{figure}[htb!]
	\centering
	\includegraphics[scale=1.0]{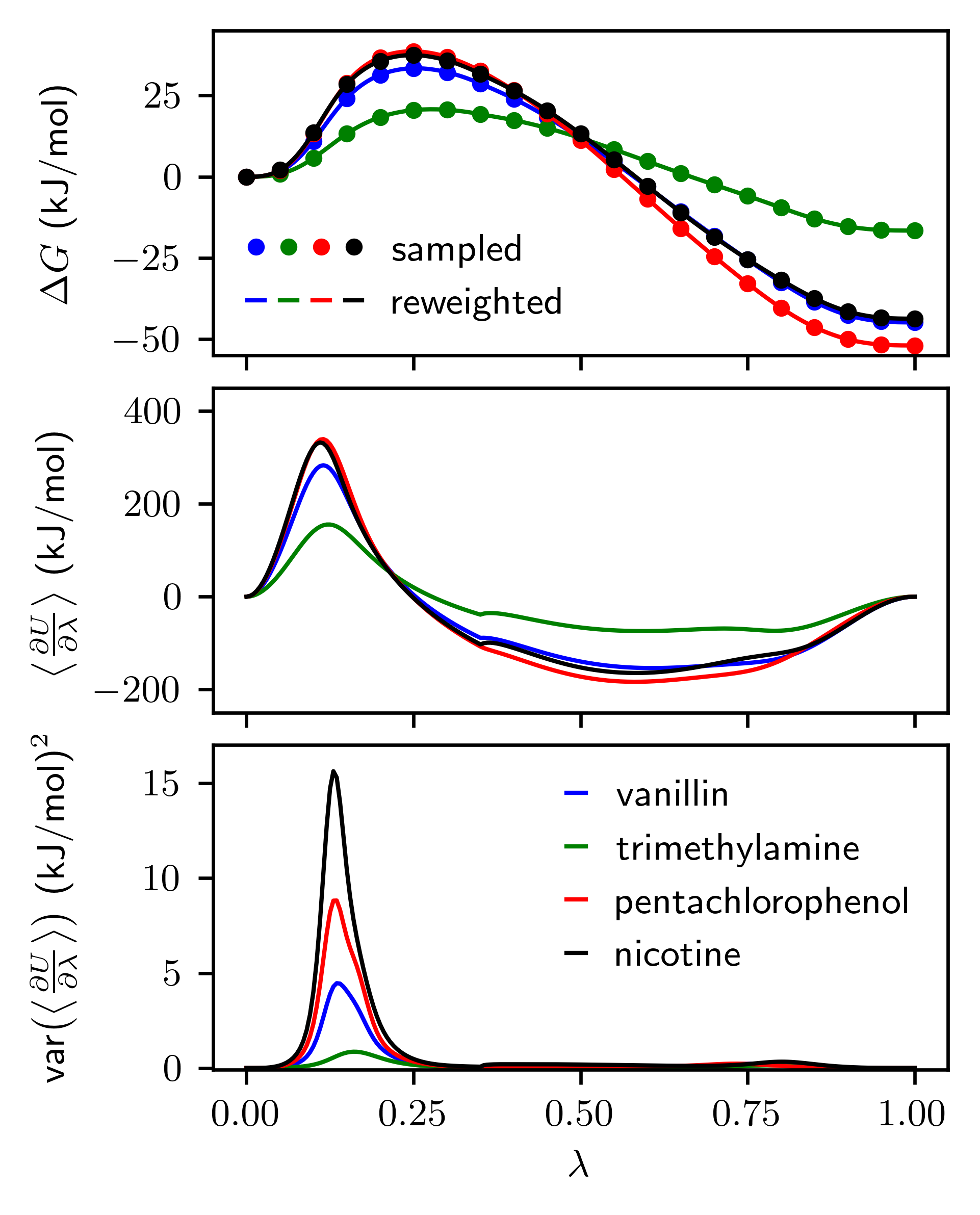}
	\caption{
    Property profiles for the solvation of multiple solutes in 1-octanol obtained by a concerted pathway with smooth switching functions with consensus parameter values.
	Circles and solid lines indicate simulation and reweighting results, respectively.
	}
	\label{fig:octanol_energy_total}
\end{figure}

Figure~\ref{fig:octanol_energy_total} allows an in-depth analysis of the concerted alchemical pathway for the different solutes in 1-octanol.
The free energy exhibits smooth variations, the $\langle \partial U/ \partial \lambda \rangle$ curve is devoid of narrow valleys, and the variance is in a proper order of magnitude.
The same conclusions are valid for the water medium results shown in the Supporting Information.
We can observe minor kinks in the mean derivative curves at $\lambda = 0.35$, where the coupling of the residual and electrostatic potentials starts (i.e., $\lambda = \lambda_\textsc{r}^\text{s} = \lambda_\textsc{e}^\text{s}$).
Hence, the second-order derivative of the free energy profile is discontinuous at such a point.
This is not surprising because the employed switching functions, defined in eqs~\eqref{eq:sfnova} and \eqref{eq:SFsmooth}, are second-order differentiable only at both $\lambda_k^\text{s}$ and $\lambda_k^\text{f}$.
Nevertheless, the free energy profile is continuous everywhere, and the presence of kinks in its derivative does not influence the free energy calculation via independent simulations of intermediate states.
However, in the context of $\lambda$ dynamic simulations, where the free energy derivative is the mean force exerted on $\lambda$, it might be advisable to use an even smoother family of sigmoid-shaped switching functions.

In the Supporting Information, we provide figures aimed at studying simulations performed using the sequential LBF approach with quartic switching functions for the four solutes. They show the negligible contribution of the residual potential to the total variance of the alchemical transformation concerning the other potentials. It is interesting to note that despite this, the residual potential contributes significantly to the free energy. We can also verify that the free energy variances with water as the solvent were smaller than those with 1-octanol as the solvent. In the latter case, the electrostatic potential had a significant weight in the total variance, which was not observed with the water system.

Table~\ref{tab:partition_coefficient_results} shows the logarithms of partition coefficients calculated from the free energy data presented in Table~\ref{tab:energy_total}. 
Experimental data from the literature \cite{Leo_1971,Hansch_1995} are also provided for comparison.
All of the values are positive, which indicates that all of the solutes have a greater affinity for 1-octanol than for water.
For each chemical species, we can observe good agreement among the results of simulations using the soft-core model and the LBF schemes but significant deviations in comparison with the experimental data.
For example, the deviations are around 120\% for nicotine and 37\% for vanillin.
This issue is due to the inadequacy of the employed force fields rather than the calculation methods.

\begin{table*}[htbp!]
	\centering
	\setlength\tabcolsep{0pt}
	\caption{
    Logarithms of the octanol-water partition coefficients for different solutes obtained using the free energy data in Table~\ref{tab:energy_total}; data found in the literature are also reported.
	}
	\label{tab:partition_coefficient_results}
	\begin{tabular*}{\textwidth}{@{\extracolsep{\fill}}lccccc}
		\hline
		Solute & Soft-core & Concerted LBF & Sequential LBF & Ref.~\citenum{Bannan_2016} & Experimental \\ 		
		\hline
        TMA & $0.86 \pm 0.06$  & $0.94 \pm 0.05$ & $0.96 \pm 0.06$ & $1.09 \pm 0.03$ & $0.27$ \cite{Leo_1971} \\
        Nicotine & $2.51 \pm 0.08$  & $2.51 \pm 0.11$ & $2.49 \pm 0.09$ & $3.19 \pm 0.06$ & $1.17$ \cite{Leo_1971} \\
        PCP & $6.62 \pm 0.08$ & $6.68 \pm 0.09$ & $6.65 \pm 0.08$ & $6.53 \pm 0.05$ & $5.01$ \cite{Leo_1971}\\
		Vanillin & $1.52 \pm 0.10$  & $1.66 \pm 0.08$ & $1.66 \pm 0.09$ &  $1.70 \pm 0.05$ & $1.21$ \cite{Hansch_1995}  \\ 
		\hline
		\end{tabular*}
\end{table*}

\section{Conclusions}
\label{sec:conclusion}

We have proposed a new splitting of the Lennard-Jones (LJ) potential into capped and residual terms and employed the linear basis function (LBF) formalism to perform alchemical free energy calculations.
This splitting is more straightforward than previous proposals because it involves coefficients independent of the LJ parameters.
Additionally, we applied a set of smooth switching functions to control the coupling of the multiple potential energy terms, allowing a concerted evaluation of LJ and Coulomb components of the solvation free energy.
The capped potential avoids numerical issues by replacing the native LJ potential at the onset of the alchemical transformation.
The coupling of the residual and electrostatic potentials starts at a later stage when the risk of numerical instabilities has vanished.
The method facilitates code development, simulation planning, and postprocessing activities while being on par with other methods in terms of efficiency.

The inherent reweighting-related features of the LBF model allowed us to carry out fast optimization procedures.
As a result, we could find switching function parameters that led to safe and efficient alchemical pathways for solutes spanning a wide range of molecular sizes and hydration free energies.
Given the similarity of the optimal parameters, we determined consensus values that might be suitable for general solvation free energy calculations.
If this is not the case for a specific system, using the optimization procedure presented in this work can provide an adequate set of parameters.

Using our concerted LBF approach with consensus parameters, we performed solvation free energy simulations of solutes other than those used for optimization in both water and 1-octanol media.
Their results allowed us to calculate the octanol-water partition coefficients and compare the obtained values with literature data.
We also employed the sequential LBF pathway and the ubiquitous soft-core model to perform analogous simulations.
Through a successful comparison with these established methodologies, we attested to the correctness and robustness of the concerted LBF pathway.

In addition, we observed that a reaction field model with infinite dielectric constant was the best among several pairwise options for substituting lattice-sum methods in solute-solvent electrostatic calculations.

\begin{acknowledgement}
	The authors acknowledge the financial support from FAPERJ, CNPq, CAPES, and ANP-PETROBRAS. We also thank the ``Laboratório Nacional de Computação Científica'' (LNCC/MCTI, Brazil) for the use of the \href{https://sdumont.lncc.br}{SDumont} supercomputer.
\end{acknowledgement}

\begin{suppinfo} 
Additional results from optimization with scheme Q applying quartic switching functions;
Lennard-Jones and electrostatic contributions of solvation free energies obtained from soft-core-based simulations;
additional property profiles computed by the concerted and sequential LBF pathways with smooth switching functions;
efficiency comparison through graphs of uncertainties \textit{vs}. total simulated times;
and details on the implementation of the method in the OpenMM software package.
\end{suppinfo}

\bibliography{main}

\end{document}


\thispagestyle{empty}

\maketitle

\newpage

\section{Supplementary Figures and Tables}

\begin{figure}[htb!]
	\centering
	\includegraphics[scale=1.0]{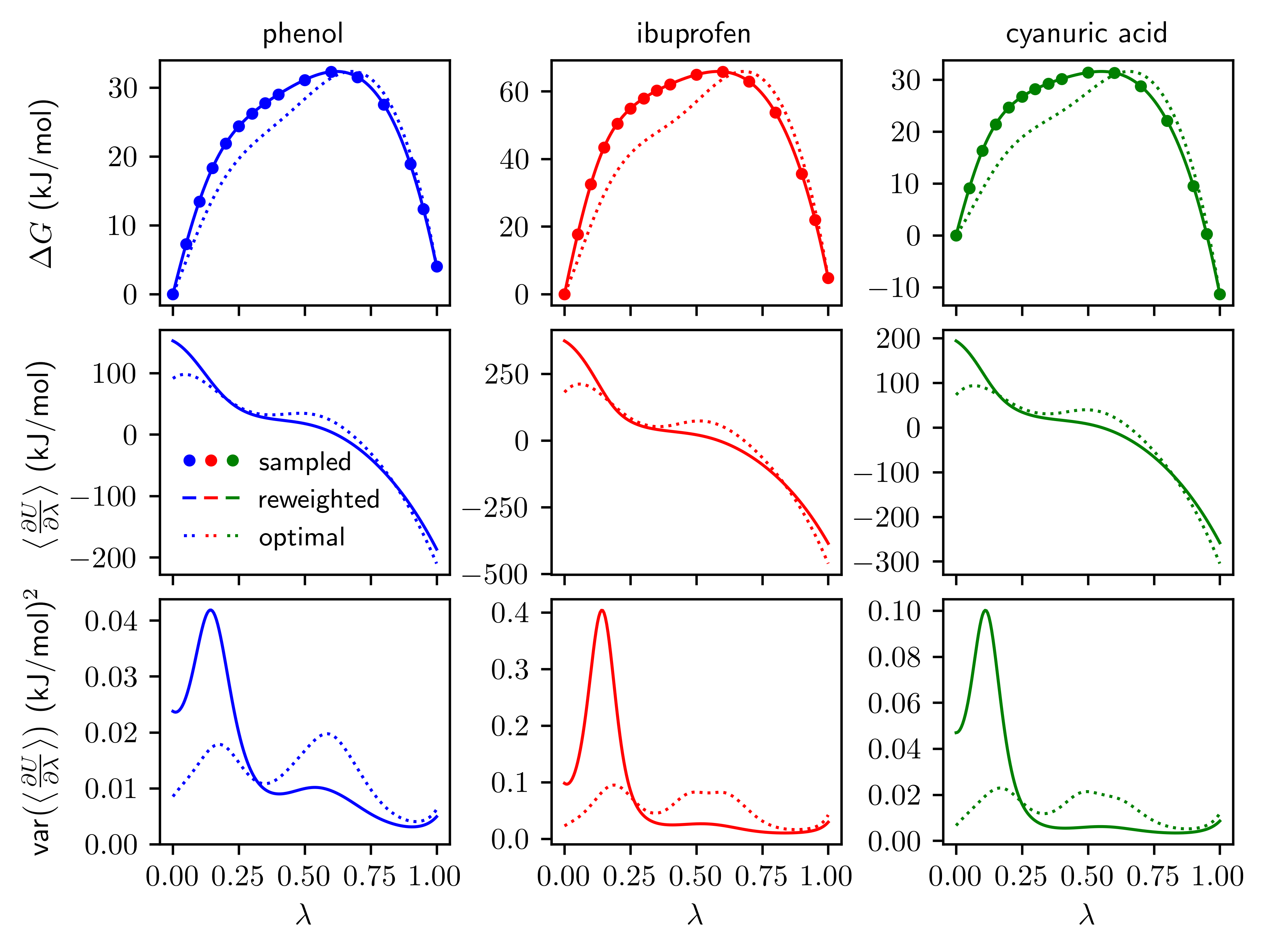}
	\caption{
	Property profiles regarding the coupling of the capped-LJ potential with a quartic switching function in the hydration of phenol, ibuprofen, and cyanuric acid. Circles and solid lines respectively indicate simulation and reweighting results for the reference system. Dashed lines represent reweighting results in the case of the optimal switching function derived from Scheme Q.
	}
	\label{fig:cap2}
\end{figure}

\begin{figure}[htb!]
	\centering
	\includegraphics[scale=1.0]{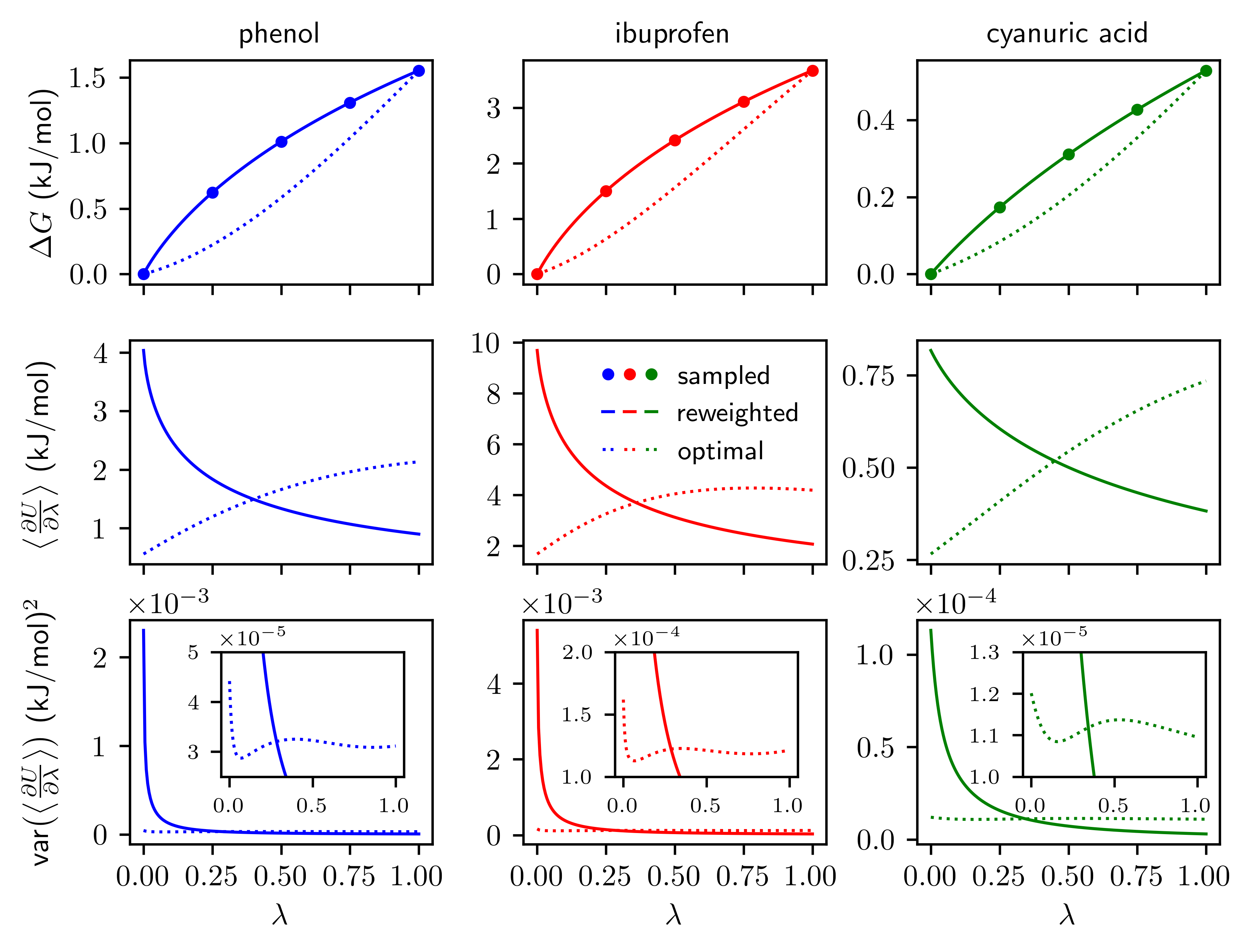}
	\caption{
	Property profiles regarding the coupling of the residual-LJ potential with a quartic switching function in the hydration of phenol, ibuprofen, and cyanuric acid. Circles and solid lines respectively indicate simulation and reweighting results for the reference system. Dashed lines represent reweighting results in the case of the optimal switching function derived from Scheme Q.
	}
	\label{fig:decap}
\end{figure}

\begin{figure}[htb!]
	\centering
	\includegraphics[scale=1.0]{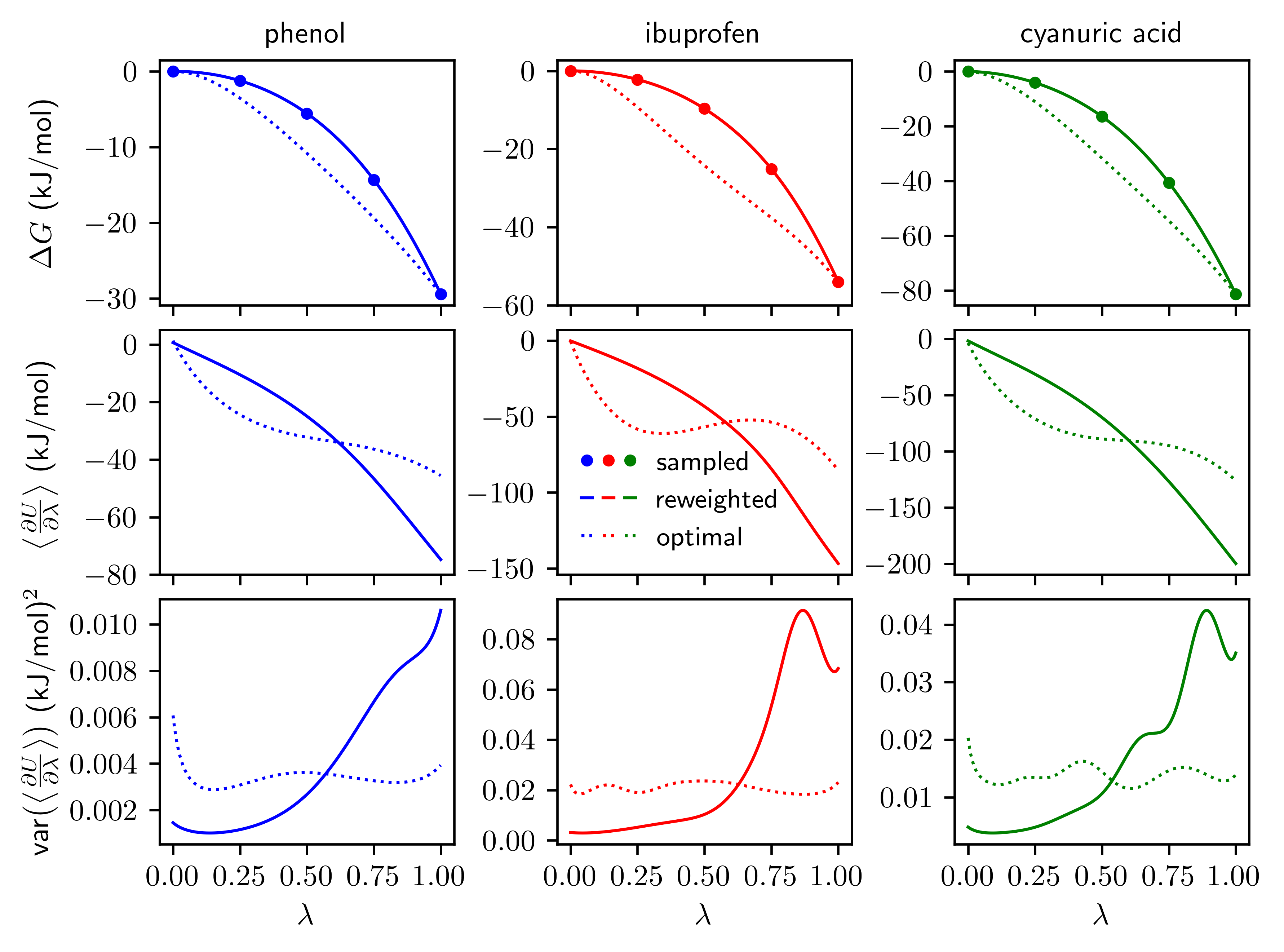}
	\caption{
	Property profiles regarding the coupling of the electrostatic potential with a quartic switching function in the hydration of phenol, ibuprofen, and cyanuric acid.
	Circles and solid lines respectively indicate simulation and reweighting results for the reference system. Dashed lines represent reweighting results in the case of the optimal switching function derived from Scheme Q.
	}
	\label{fig:elec}
\end{figure}

\begin{figure}[htb!]
	\centering
	\includegraphics[scale=1.0]{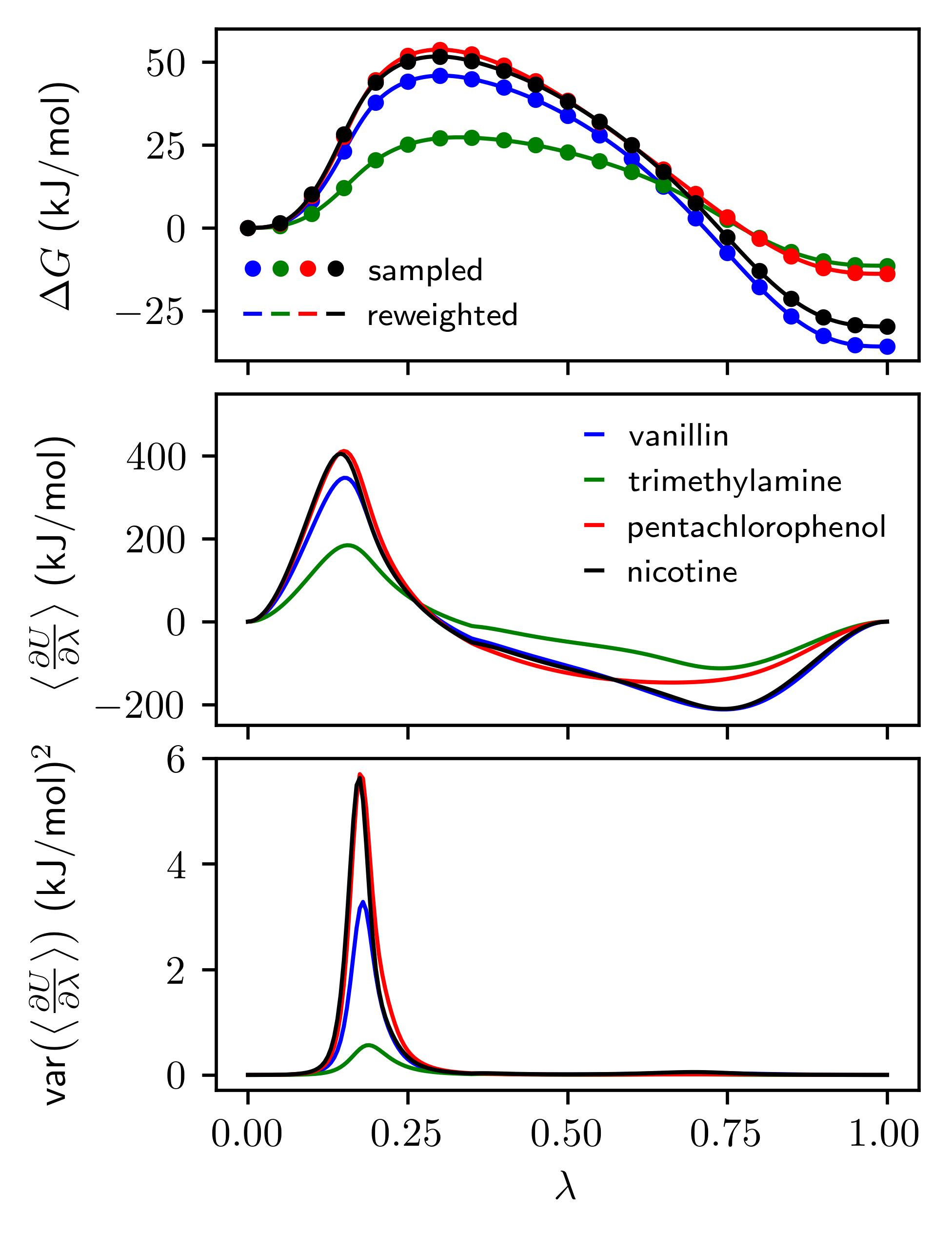}
	\caption{Property profiles for the solvation of multiple solutes in water, obtained by a concerted LBF pathway with smooth switching functions and consensus parameters. Circles and solid lines indicate simulation and reweighting results, respectively.}
	\label{fig:water_energy_total}
\end{figure}

\begin{figure}[htb!]
	\centering
	\includegraphics[scale=1.0]{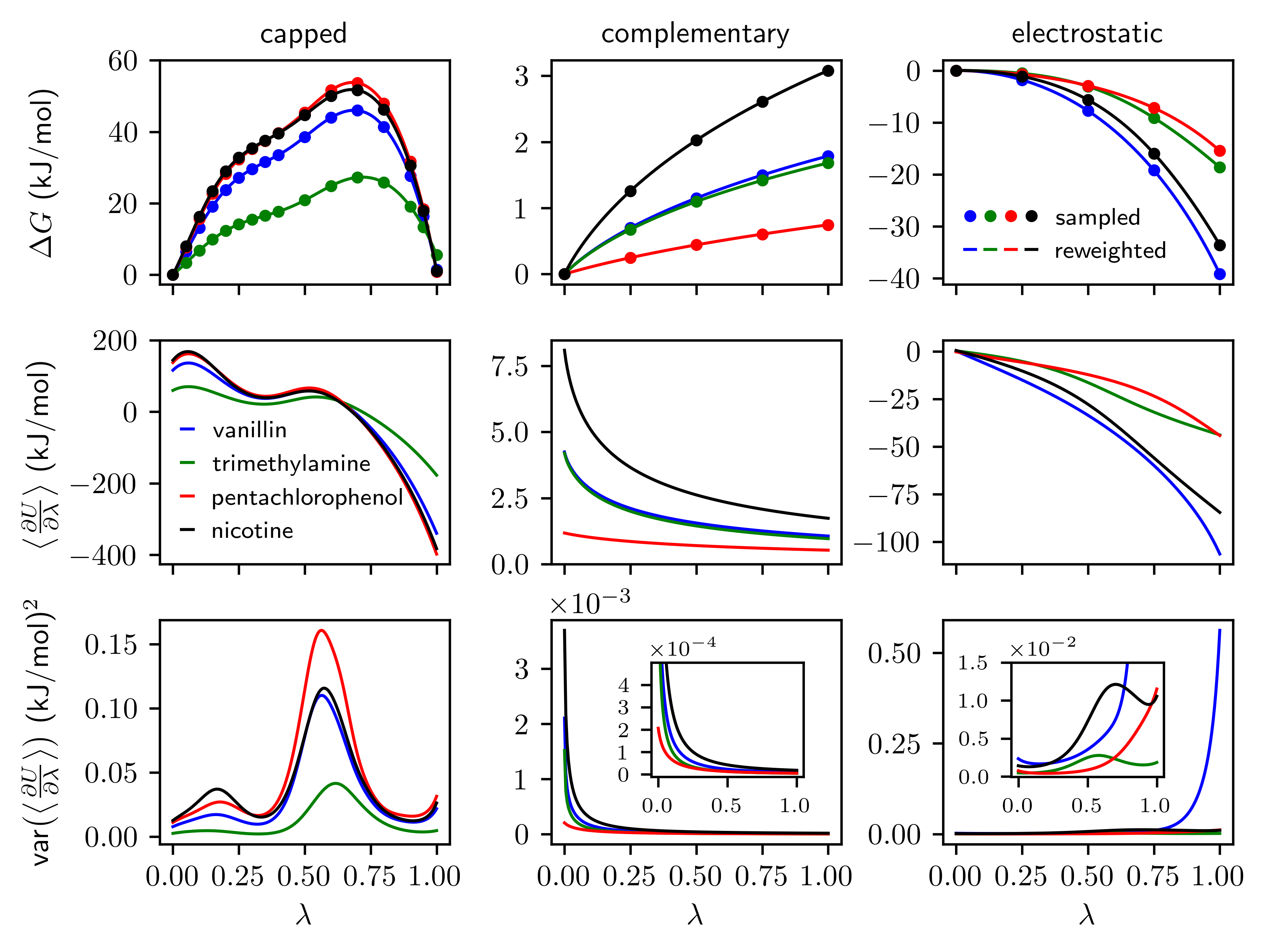}
	\caption{Property profiles for the solvation of multiple solutes in water, obtained by a sequential LBF pathway with quartic switching functions and consensus parameters. Circles and solid lines indicate simulation and reweighting results, respectively.}
	\label{fig:water_energy_total_III}
\end{figure}

\begin{figure}[htb!]
	\centering
	\includegraphics[scale=1.0]{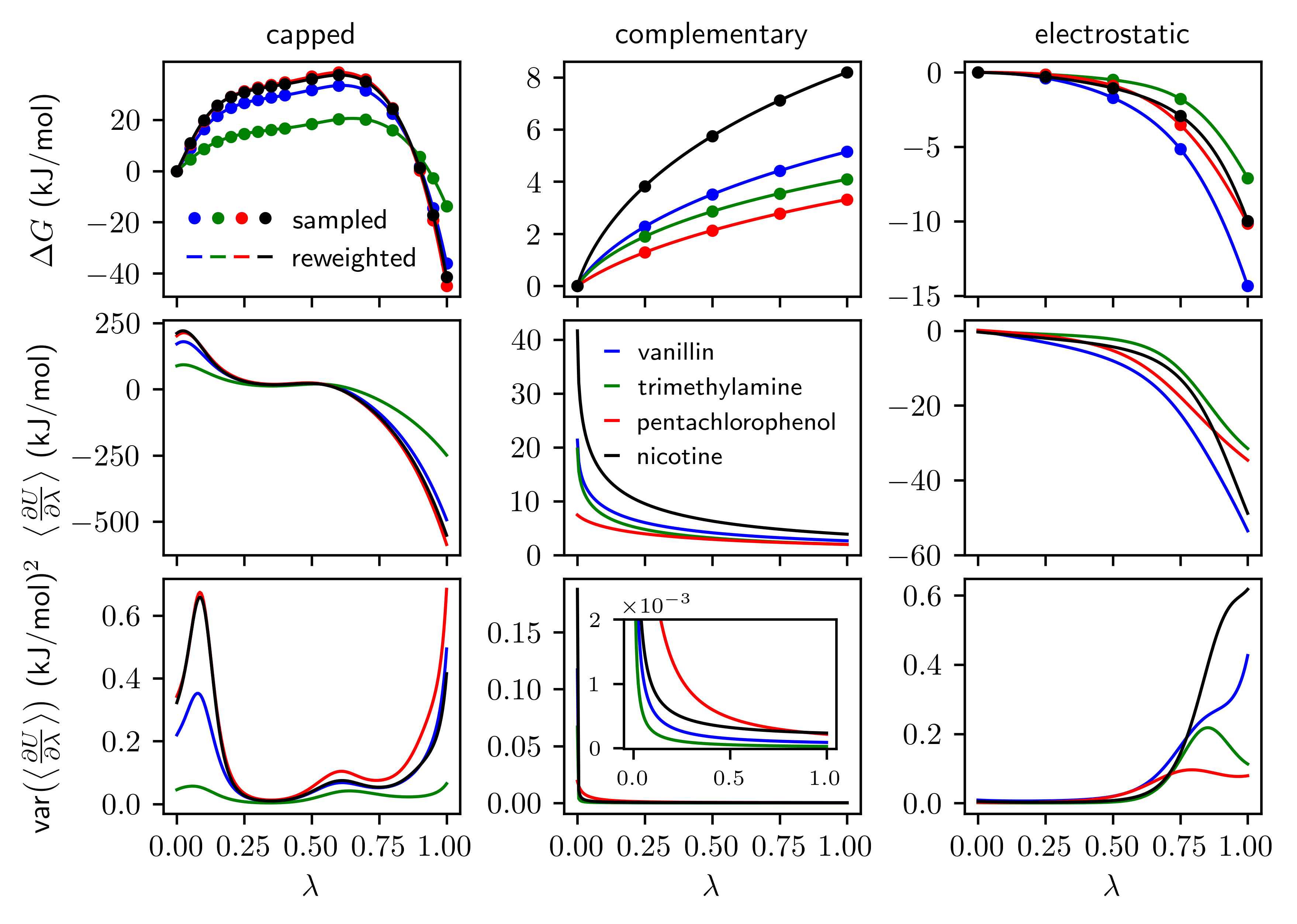}
	\caption{Property profiles for the solvation of multiple solutes in 1-octanol, obtained by a sequential LBF pathway with quartic switching functions and consensus parameters. Circles and solid lines indicate simulation and reweighting results, respectively.}
	\label{fig:octanol_energy_total_III}
\end{figure}

\clearpage

\begin{figure}[htb!]
	\centering
	\includegraphics[scale=1.0]{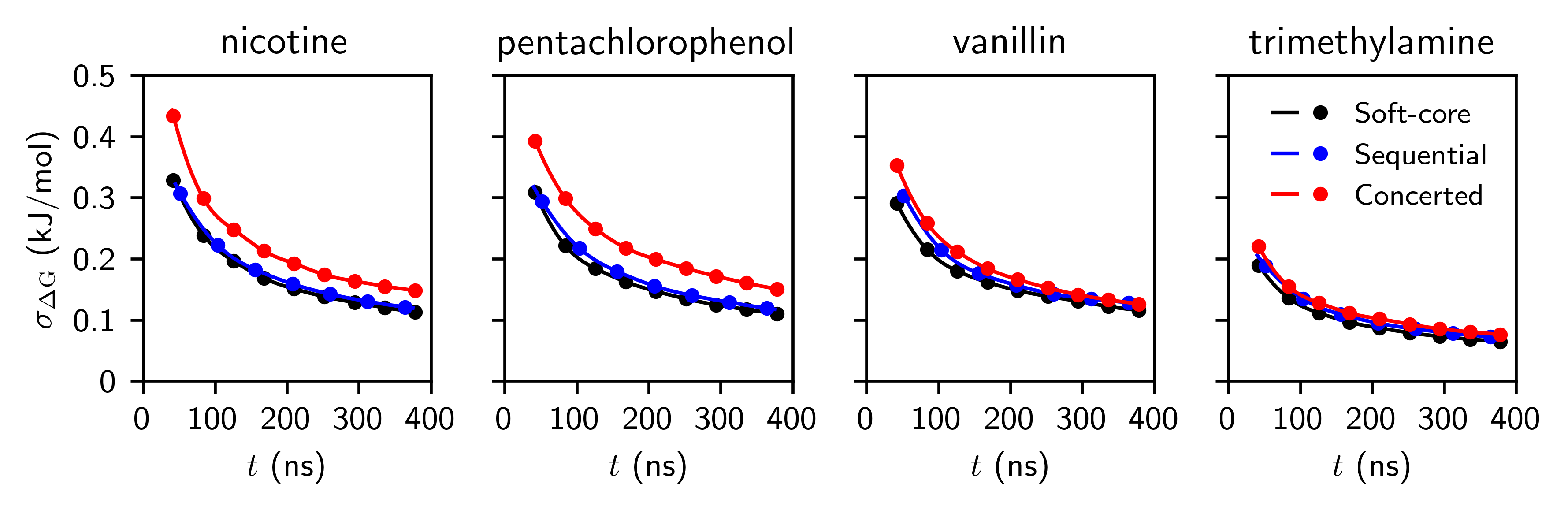}
	\caption{Mean-square errors of solvation free energies estimated via MBAR for different solutes in water as a function of the total simulated time (i.e. the sum for all $\lambda$ states). The values obtained by the soft-core model as well as the sequential and concerted LBF alchemical pathways are compared. Circles and solid lines indicate simulation and interpolated results, respectively.
	The concerted LBF method yields uncertainties only slightly larger than the soft-core and sequential LBF approaches.	
}
	\label{fig:stdev_water}
\end{figure}

\begin{figure}[htb!]
	\centering
	\includegraphics[scale=1.0]{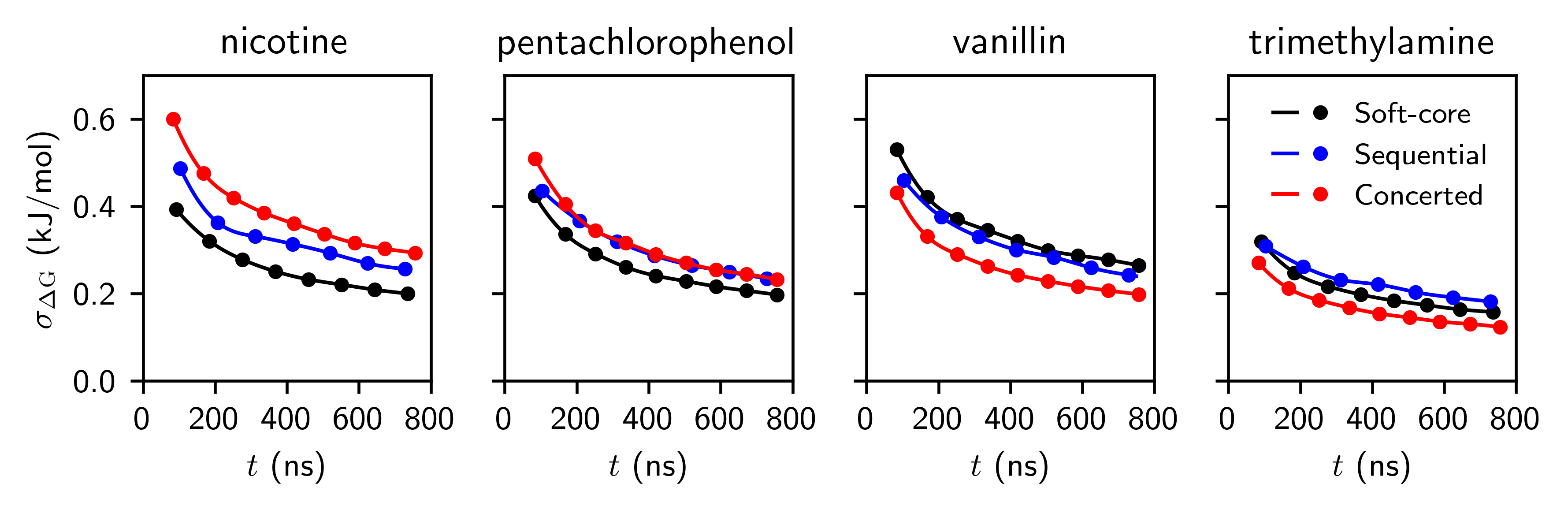}
	\caption{Mean-square errors of solvation free energies estimated via MBAR for different solutes in 1-octanol as a function of the total simulation time (i.e. the sum for all $\lambda$ states). The values obtained by the soft-core model as well as the sequential and concerted LBF alchemical pathways are compared. Circles and solid lines indicate simulation and interpolated results, respectively.
	The concerted LBF method is the best one when vanillin or trimethylamine is the solute.
}
	\label{fig:stdev_octanol}
\end{figure}

\begin{table*}[htb!]
	\centering
	\setlength\tabcolsep{0pt}
	\caption{Free energy values in kJ/mol for the coupling of Lennard-Jones ($\Delta G_\textsc{lj}$) and electrostatic ($\Delta G_\textsc{e}$) potentials in the application of the concerted LBF, sequential LBF, and soft-core alchemical pathways. Note: a single electrostatic contribution calculation applies for both the sequential and soft-core pathways.}
	\label{tab:schemeIIIa}
	\begin{tabular*}{\textwidth}{@{\extracolsep{\fill}}lccccc}
		\hline
		\multirow{2}{*}{Solute} & \multirow{2}{*}{Pathway} & \multicolumn{2}{c}{Water as solvent} & \multicolumn{2}{c}{1-Octanol as solvent} \\
		\cline{3-4}\cline{5-6}
		& 
		& $\Delta G_\textsc{lj}$
		& $\Delta G_\textsc{e}$
		& $\Delta G_\textsc{lj}$
		& $\Delta G_\textsc{e}$
		\\ 		
		\hline
		\multirow{3}{*}{trimethylamine} & Concerted & $7.2 \pm 0.1$ & $-18.5 \pm 0.1$ & $-9.3 \pm 0.2$ & $-7.3 \pm 0.2$ \\
		& Sequential & $7.2 \pm 0.1$ & \multirow{2}{*}{$-18.46 \pm 0.06$} & $-9.6 \pm 0.2$ & \multirow{2}{*}{$-7.1 \pm 0.3$} \\
		& Soft-core & $7.1 \pm 0.1$ & & $-9.3 \pm 0.2$ & \\
		\hline
		\multirow{3}{*}{nicotine} & Concerted & $3.9 \pm 0.2$ & $-33.3 \pm 0.2$  & $-33.8 \pm 0.5$ & $-9.9 \pm 0.4$ \\
		& Sequential & $4.1 \pm 0.1$ & \multirow{2}{*}{$-33.1 \pm 0.1$} & $-33.2 \pm 0.3$ & \multirow{2}{*}{$-10.0 \pm 0.4$} \\
		& Soft-core & $4.0 \pm 0.2$ & & $-33.4 \pm 0.3$ & \\
		\hline
		\multirow{3}{*}{pentachlorophenol} & Concerted & $1.5 \pm 0.2$ & $-15.4 \pm 0.2$ & $-41.7 \pm 0.4$ & $-10.3 \pm 0.3$ \\
		& Sequential & $1.6 \pm 0.2$ & \multirow{2}{*}{$-15.42 \pm 0.06$} & $-41.6 \pm 0.3$ & \multirow{2}{*}{$-10.2 \pm 0.3$} \\
		& Soft-core & $1.6 \pm 0.2$ & & $-41.4 \pm 0.3$ & \\
		\hline
		\multirow{3}{*}{vanillin} & Concerted & $3.2 \pm 0.2$ & $-38.7 \pm 0.2$ & $-31.2 \pm 0.3$ & $-13.8 \pm 0.3$ \\
		& Sequential & $3.2 \pm 0.2$ & \multirow{2}{*}{$-39.0 \pm 0.2$} & $-31.0 \pm 0.3$ & \multirow{2}{*}{$-14.3 \pm 0.4$} \\
		& Soft-core & $3.2 \pm 0.2$ & & $-30.2 \pm 0.3$ & \\ 
		\hline& & 
	\end{tabular*}
\end{table*}

\begin{table*}[htb!]
	\centering
	\setlength\tabcolsep{0pt}
	\caption{Free energy values in kJ/mol for the coupling of capped ($\Delta G_\textsc{c}$) and residual ($\Delta G_\textsc{r}$) potentials in the application of sequential and concerted LBF alchemical pathways. The concerted values were calculated considering the milestone states of the sequential pathway.}
	\label{tab:schemeIIIb}
	\begin{tabular*}{\textwidth}{@{\extracolsep{\fill}}llcccc}
		\hline
		\multirow{2}{*}{Solute} & \multirow{2}{*}{Pathway} & \multicolumn{2}{c}{Water as solvent} & \multicolumn{2}{c}{1-Octanol as solvent} \\
		\cline{3-4}\cline{5-6}
		&
		& $\Delta G_\textsc{c}$
		& $\Delta G_\textsc{r}$
		& $\Delta G_\textsc{c}$
		& $\Delta G_\textsc{r}$ 
		\\ 		
		\hline
		\multirow{2}{*}{trimethylamine} & Concerted & $5.56 \pm 0.07$ & $1.7 \pm 0.1$ & $-13.4 \pm 0.1$ & $4.2 \pm 0.2$ \\
		& Sequential & $5.5 \pm 0.1$ & $1.682 \pm 0.009$ & $-13.7 \pm 0.2$ & $4.09 \pm 0.02$ \\
        \hline
		\multirow{2}{*}{nicotine} & Concerted  & $0.9 \pm 0.1$ & $3.1 \pm 0.2$ & $-42.1 \pm 0.3$ & $8.3 \pm 0.4$ \\
		& Sequential & $1.0 \pm 0.2$ & $3.08 \pm 0.02$ & $-41.4 \pm 0.3$ & $8.19 \pm 0.04$ \\
        \hline
		\multirow{2}{*}{pentachlorophenol} & Concerted & $0.8 \pm 0.1$ & $0.7 \pm 0.2$ & $-45.0 \pm 0.2$ & $3.3 \pm 0.3$ \\
		& Sequential & $0.8 \pm 0.2$ & $0.745 \pm 0.008$ & $-44.9 \pm 0.3$ & $3.32 \pm 0.05$ \\
        \hline
		\multirow{2}{*}{vanillin} & Concerted & $1.4 \pm 0.1$ & $1.8 \pm 0.2$  & $-36.4 \pm 0.2$ & $5.2 \pm 0.2$ \\
		& Sequential & $1.4 \pm 0.2$ & $1.79 \pm 0.01$ & $-36.1 \pm 0.3$ & $5.15 \pm 0.03$ \\
		\hline
	\end{tabular*}
\end{table*}

\clearpage

\section{Implementation Details}

As mentioned in the main article, all molecular dynamics simulations were performed with OpenMM (version 7.5.0).
We implemented both Lennard-Jones (LJ) and reaction field terms of the solute-solvent (AS) interactions by using the \texttt{addInteractionGroup} method of the \texttt{CustomNonbondedForce} class, thus applying them only for pairs formed by a solute atom and a solvent one.
Regarding the solute-solute and solvent-solvent (AA+SS) non-bonded interactions, we use OpenMM's standard \texttt{NonbondedForce} class.
Since the solute molecule comprises a tiny fraction of the atoms in the system, we treat as exceptions all non-bonded interactions internal to the solute, just like one usually treats 1-4 interactions.
At the same time, we set the electric charges and LJ epsilon parameters of solute atoms to zero, turning off all solute-solvent interactions.
Such a measure also turns off the electrostatic interaction between the solute molecule and its periodic images, which is consistent with the reaction-field nature of the solute's interaction with its surrounding.
Compared to an unmodified \texttt{NonbondedForce}, the drop in performance of the described procedure is minimal.


We apply the PME method at $\lambda = 1$ to compute the difference $U_\textsc{pme}^\textsc{as}-U_\textsc{crf}^\textsc{as}$ and, eventually, the free-energy correction in Eq.~(15) of the main article.
For this, we resort to the \texttt{addParticleParameterOffset} method of the \texttt{NonbondedForce} class, which allows us to scale the solute charges by a user-defined prefactor.
During the simulation, we simply set this prefactor to zero.
At regular intervals, we switch its value to one and back to zero, computing the system's potential energy at both conditions, whose difference yields $U_\textsc{pme}^\textsc{as}$.
Because the correlation times are usually large, the sampling frequency can be small and, as a result, the described procedure does not cause much overhead.

A repository containing python scripts applied in our solvation free energy simulations and post-processing activities is available at \href{https://github.com/atoms-ufrj/jctc-lbf-paper}{github.com/atoms-ufrj/jctc-lbf-paper}.